\documentclass[
superscriptaddress,
twocolumn,
 amsmath,
 amssymb,
aps,
]{revtex4-2}
\usepackage{import}
\usepackage{lipsum}
\usepackage{braket}
\usepackage{graphicx}
\usepackage{dcolumn}
\usepackage{siunitx}
\usepackage[normalem]{ulem}
\usepackage{upgreek}
\usepackage{bm}
\usepackage{color}
\usepackage{notes2bib}
\usepackage{pgf}
\usepackage{natbib}
\usepackage{hyperref}
\usepackage{svg}
\usepackage[utf8]{inputenc}
\DeclareUnicodeCharacter{2212}{-}

\definecolor{d_Red}{RGB}{190, 30, 45}
\definecolor{d_Cyan}{RGB}{0, 174, 239}
\definecolor{d_Gold}{RGB}{175, 169, 97}
\definecolor{d_Yellow}{RGB}{255, 213, 58}
\definecolor{d_Purple}{RGB}{104, 36, 109}
\definecolor{d_Heather}{RGB}{203,168,177}
\definecolor{d_Stone}{RGB}{218,205,162}
\definecolor{d_Sky}{RGB}{165,200,208}
\definecolor{d_Cedar}{RGB}{182,170,167}
\definecolor{d_Concrete}{RGB}{179,189,177}
\definecolor{d_Ink}{RGB}{0,42,65}
\definecolor{d_Black}{RGB}{51,49,50 }

\definecolor{palatinate}{HTML}{A300A3}
\definecolor{teal}{HTML}{00A3A3}
\definecolor{gold}{HTML}{A3A300}

\hypersetup{colorlinks=true,linkcolor=palatinate,citecolor=teal,filecolor=magenta,urlcolor=gold}
\renewcommand{\sout}[1]{}

\newcommand{\fulldiff}[2]{\frac{\mathrm{d} #1}{\mathrm{d} #2}}
\newcommand{\bracket}[2]{\langle #1|#2 \rangle}

\newcommand{\ham}{\mathrm{\hat{H}}}

\newcommand{\transition}[2]{$\ket{#1} \rightarrow \ket{#2}$}
\newcommand{\decay}[0]{\Gamma_{10}}

\begin{document}


\title{An Intuitive Visualisation Method for Arbitrary Qutrit (Three Level) States}

\author{Max Z. Festenstein}%
\affiliation{Department of Physics, Rochester Building, Durham, DH1 3LE, UK.}

\date{April 25th 2023}

\begin{abstract}
Visual methods are of great utility in understanding and interpreting quantum mechanics at all levels of understanding. The Bloch sphere, for example, is an invaluable and widely used tool for visualising quantum dynamics of a two level qubit system. In this work we present an `octant' visualisation method for qutrits bearing similarity to the Bloch sphere, that encompasses all eight degrees of freedom necessary to fully describe a three level state whilst remaining intuitive to interpret. Using this framework, a set of typical three level processes are modelled, described and displayed.
\end{abstract}

\maketitle


\section{Introduction}
In equivalence to classical computing, Quantum Information Processing focuses primarily on the dynamics of two level systems. Unlike classical computation, however, the quantum bit (qubit) exists not in a discrete space of $\{0,1 \} \subset \mathbb{Z} $ but continuously as a two level wavefunction $\ket{\psi} = \alpha \ket{0} + \beta \mathrm{e}^{-i\phi_1} \ket{1}$ in $\mathbb{C}^2$ i.e. existing as a complex, continuously varying object rather than a binary real one. This change in properties allows for significant computational speedup for certain tasks such as the seminal Shor's and Grover's algorithms for prime number factorisation \cite{Shor1994AlgorithmsFactoring} and unstructured searches \cite{Grover1996ASearch}, as well as more recent quantum machine learning \cite{Biamonte2017QuantumLearning}. Despite the prevalence of two state dynamics, three states are a key feature in many quantum systems. Examples include, Raman transition \cite{Grigoryan2001AdiabaticDetuning,Boradjiev2010StimulatedResonance,Miroshnychenko2010CoherentState}, STImulated Raman Adiabatic Passage (STIRAP) \cite{Cubel2005CoherentStates,Weidt2016Trapped-IonFields,Molony2014CreationState,Menchon-Enrich2016SpatialProgress}, Electromagnetically Induced Transparency (EIT) \cite{Fleischhauer2005ElectromagneticallyMedia,Schraft2016StoppedCrystal,Nicolas2014AQubits,Lin2016AnFunctions} and non-linear processes such as frequency doubling \cite{Kleinman1962TheoryLight,Lin1999MechanismCrystals,Cong2012Experimental3OH}  and Four Wave Mixing (FWM) \cite{Lee2016HighlySystem,Maxwell2013StoragePolaritons,Zugenmaier2018Long-livedTemperature,Willis2009Four-waveVapor,Lee2017Single-photonEnsemble,Park2019Polarization-entangledInterferometer,Noh2021Four-waveStudy}. 
\\
\\
Visualisations of quantum systems are highly useful tools to ground abstract dynamics, both for educational purposes when seeing a problem for the first time, or as a framework to illustrate novel results to others. Techniques, such as the Bloch sphere, to visualise two-state processes are well established \cite{Nielsen2010QuantumInformation} but extending beyond this, there  is growing interest in the visualisation of quantum circuits. Tools such as ZX-calculus allow for clear diagrammatic illustrations of complex processes \cite{Coecke2011InteractingDiagrammatics,Kissinger2019UniversalMeasurements,deBeaudrap2020TheSurgery,Kissinger2022SimulatingDecompositions}. Furthermore, with a multi-disciplinary convergence around the development of quantum technologies, visualisation methods can greatly aid in the development of `quantum literacy' for those without a background in the field \cite{Nita2023TheProblem-solving}. When considering visualising three levels, there is no widely adopted equivalent for three levels although various schemes have been tried \cite{Kurzynski2016Three-dimensionalQutrit,Jevtic2014QuantumEllipsoids}. In addition, in the field of quantum computing there is growing interest in three level systems or quantum trits (qutrits) \cite{Gokhale2019,Durt2003SecurityQutrits,Liu2021TopologicalParafermions,Duan2001Long-distanceOptics,Verresen2023EverythingModel}. As the dynamics in a qutrit have the scope to be much richer than their two level counterparts, a framework in which to visually represent a qutrit could provide a useful aid to help intuit the behaviour of a system. Despite the potential use of such a framework, constructing one is a non-trivial undertaking. 
\\
\\
In this work we build on the representation presented in \cite{Benhemou2021UniversalityAtoms} to allow for a description of an arbitrary qutrit state on an `octant' plot for use both as an educational tool and one for researchers to illustrate novel results. We start by describing pure states and consider a phase-sensitive interference process in section \ref{sec:Pure States}. Then in section \ref{sec:Mixed States} we extend to fully express an arbitrary mixed qutrit state. Using this framework, we then  model and display the common three level protocols FWM and EIT, illustrating and describing their dynamics.
\section{Pure State Description} \label{sec:Pure States}
In the 2 level case,  an arbitrary density matrix $\rho$ can be expressed as a weighted linear sum of the Pauli matrices, i.e. $\rho = \sum_j \alpha_j \mathbf{\sigma}_j$ for coefficients $\alpha_j \in \mathbb{R}$. These 3 Pauli matrices are the generator matrices for the Special Unitary group in two dimensions ($\mathrm{SU}(2)$) and, for 3 levels, the corresponding set of generator matrices for the $\mathrm{SU}(3)$ group are the eight Gell-Mann (GM) matrices \cite{Bertlmann2008BlochQudits}. The density matrix for any single particle qutrit state can then be expressed as $\rho = \sum_j \mathrm{a}_j \lambda_j$ with coefficients $\mathrm{a}_j \in \mathbb{R}$. 
\\
\\
Considering 2 levels, the $\mathrm{SU}(2)$ group has an isomorphic double-cover mapping onto the $\mathrm{SO}(3)$ group, letting each of the generator matrices be expressed along 3 orthogonal axes and the qubit state a vector in this space; i.e. a vector on or within the Bloch sphere. Thus for $\mathrm{SU}(2)$ it is straightforward to interpret the information being displayed due to the near-direct correspondence of state in $\mathbb{C}^2$ to a position in $\mathbb{R}^3$. Futhermore, this representation offers the ability to distinguish similarity between states; i.e. two similar (dissimilar) states $\ket{\psi}$ and $\ket{\phi}$, where $\bracket{\phi}{\psi} \approx 1 (0)$, will lie close (on opposite poles) to each other on the Bloch sphere. In the 3 level case, however, displaying the full parameter space in a format that is straightforward to interpret becomes a non trivial task due to the 8 independent parameters of the object being described. As such, to reduce the complexity of the initial task, pure states are first considered due to the more straightforward dynamics and reduced dimensionality.
\subsection{Theory} \label{subsec:Pure Theory}

\begin{figure}[h]
    \centering
    \includegraphics[width = \linewidth]{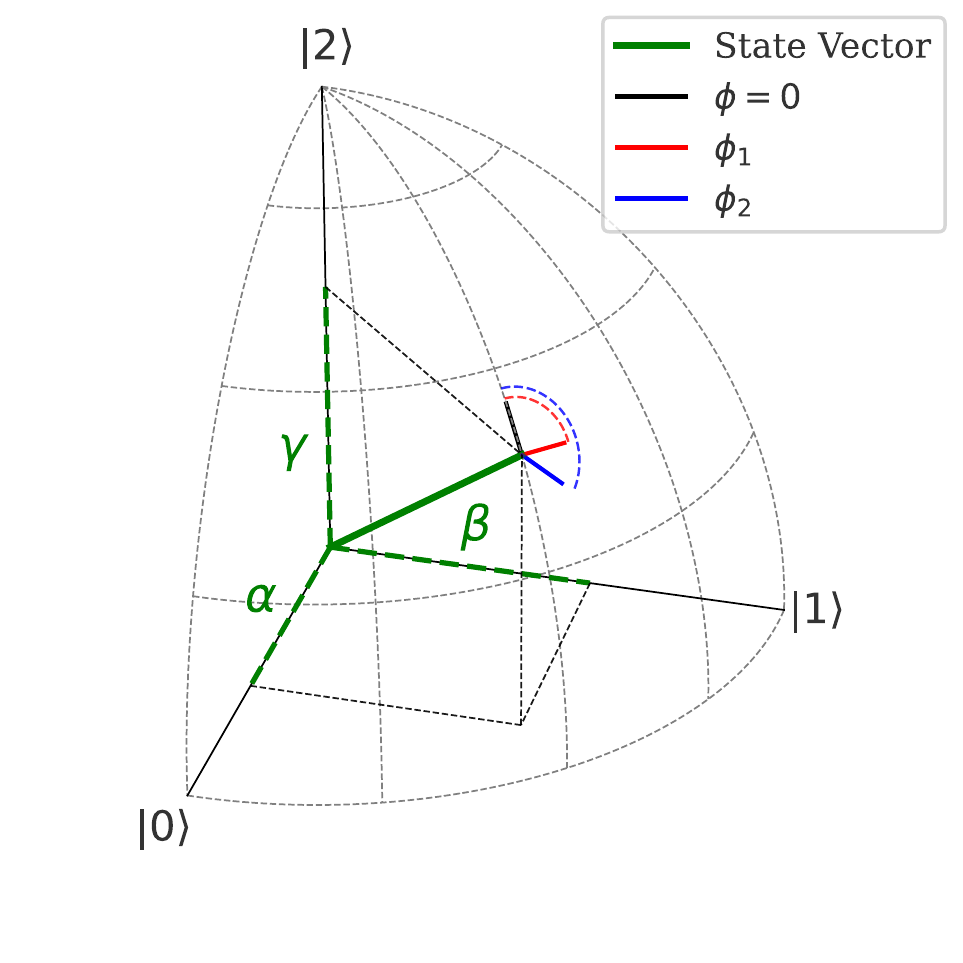}
    \caption{Graphical Representation of the qutrit state \\ $\ket{\psi} = \frac{1}{\sqrt{3}}(\ket{0} + \mathrm{e}^{-i \frac{\pi}{2}} \ket{1} + \mathrm{e}^{-i \frac{3\pi}{4}} \ket{2})$. The dashed green lines represent the magnitude of the populations in each state, with the solid green line showing the overall position of the state vector on the octant. The solid red and blue lines represent the phase values $\phi_1$ and $\phi_2$ respectively, with the coloured dashed lines guiding the eye to show how far from $\phi = 0$ (solid black) each have been rotated. In this single plot, all 5 parameters: the three populations and two phasors, can be visualised simultaneously.}
    \label{fig:pure state only}
\end{figure}

To begin with, we consider a pure state of the form

\begin{equation} \label{eq:Pure State}
\ket{\psi} = \begin{pmatrix}
   \alpha\\
    \beta \mathrm{e}^{-i\phi_1}\\
    \gamma \mathrm{e}^{-i\phi_2} \\
    \end{pmatrix}
\end{equation}
where $\alpha,\beta,\gamma \in \mathbb{R}$ and $\alpha^2 + \beta^2 + \gamma^2 = 1$. In this case, the description illustrated in \cite{Benhemou2021UniversalityAtoms} is sufficient to fully describe the system. The model used there relied on two plots side by side to show phase and state population separately. The first adjustment made here is to condense the information to a single plot by projecting the phase information as rotated lines centred at the end of the state vector, akin to hands on a clock. An example of a state represented this way is shown in figure \ref{fig:pure state only}. 

\begin{figure}[h]
    \centering
    \includegraphics[width = \linewidth]{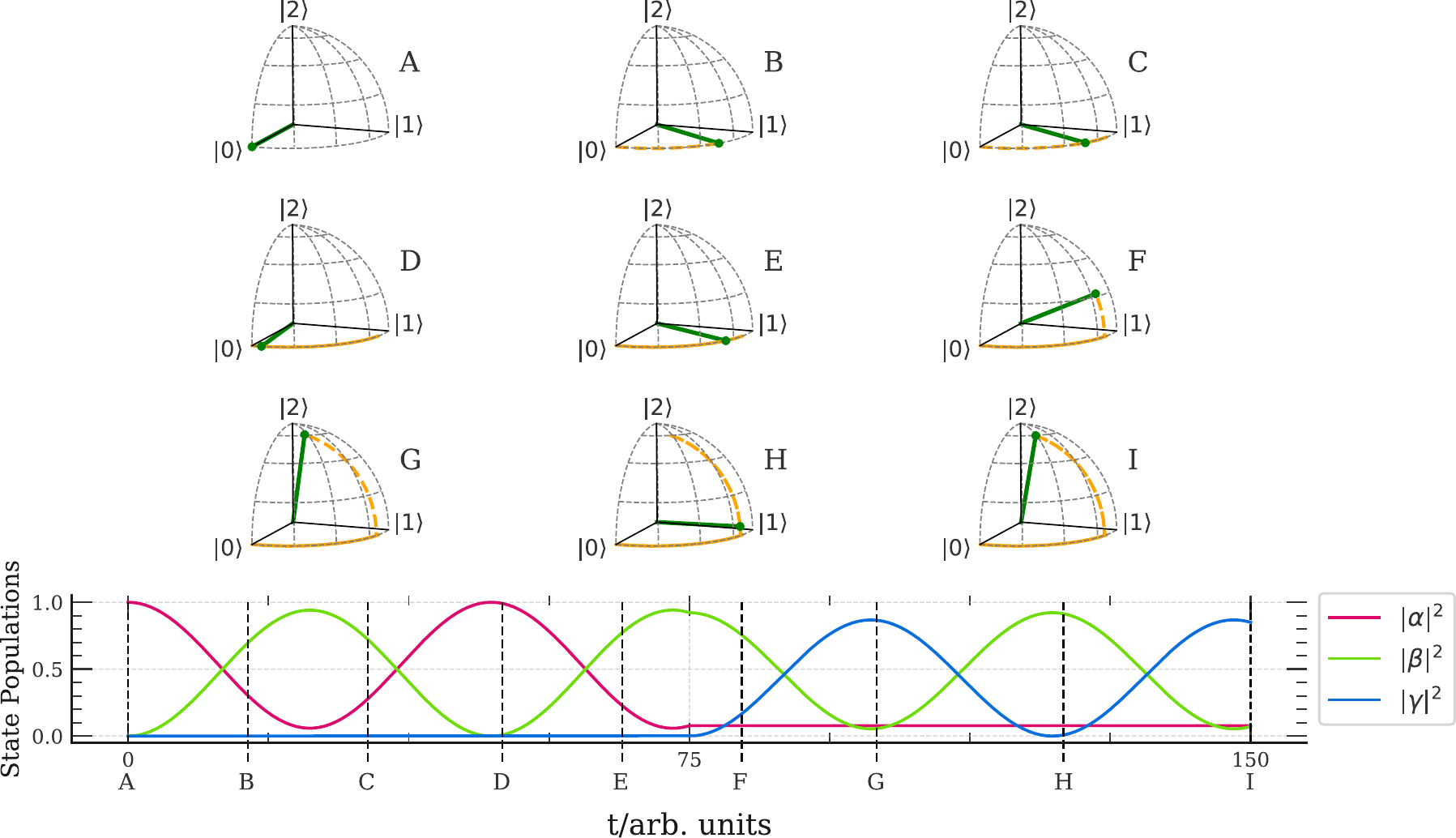}
    \caption{\textbf{Upper}: Octant plots displaying Rabi oscillations in a 3 level system. The green line and point marker display the eigenstate populations and the dashed orange line shows the path traced out by this vector. No clock hands are included here due to the lack of any oscillation or decay in the resonant case free of decay modes. \textbf{Lower}: A more traditional state population vs time evolution of the qutrit as a reference. The times where octant plots are drawn are marked with lettered black dashed lines.}
    \label{fig:Pure Oscillator Mosaic}
\end{figure}

There are two exceptions to the general requirement of needing 2 hands to represent a pure state. The first is the trivial case of being entirely in a single eigenstate where there would be no hands present at all as there is superposition where phase can accumulate; the second is when a superposition is present between only two states, resulting in only one non-vanishing coherence term and correspondingly one clock hand for a vector along any of 3 quadrants between eigenstates. To properly describe a superposition between $\ket{1}$ and $\ket{2}$, the phase difference $\phi_{12} = \phi_2 - \phi_1$ acts as this single phase term. To provide a more dynamic example of the evolution of a pure state, a test case using the Hamiltonian 
\begin{equation} \label{eq:Test Ham}
\ham = \begin{pmatrix}
   0 &\frac{\Omega_1(t)}{2}(t) & 0\\
    \frac{\Omega_1(t)}{2} & 0 & \frac{\Omega_2(t)}{2} \\
    0& \frac{\Omega_2(t)}{2} & 0 \\
    \end{pmatrix},
 \end{equation}
with time dependent Rabi frequencies $\Omega_1$ and $\Omega_2$ where $0 \leq t < 75: \Omega_1  = \frac{0.02}{2\pi}, \Omega_2 = 0 $ and $75 \leq t \leq 150: \Omega_1  = 0, \Omega_2 = \frac{0.02}{2\pi}$, is shown in figure \ref{fig:Pure Oscillator Mosaic}. In this case, oscillation of population between states appears as harmonic motion of the green state vector between the two addressed states akin to a pendulum swing, unlike the 2D case where the state precesses around the surface of the Bloch sphere.

Initially, when the transition between \transition{0}{1} is addressed (A-E) this induces Rabi oscillations between these states, which is visually represented by the state vector oscillating in the $x-y$ plane. When $t > 75$ this Rabi oscillation occurs between \transition{1}{2} (F-I), which corresponds to oscillation in the $y-z$ plane. As the population transfer to $\ket{1}$ is incomplete when this oscillation occurs, the state vector retains an $x$ axis ($\ket{0}$) component and doesn't experience a full transfer into $\ket{2}$ at the peak of oscillation.
\subsection{Two Pulse Sequence} \label{subsec:Qutrit Ramsey}
\begin{figure}[h]
    \centering
    \includegraphics[width = \linewidth]{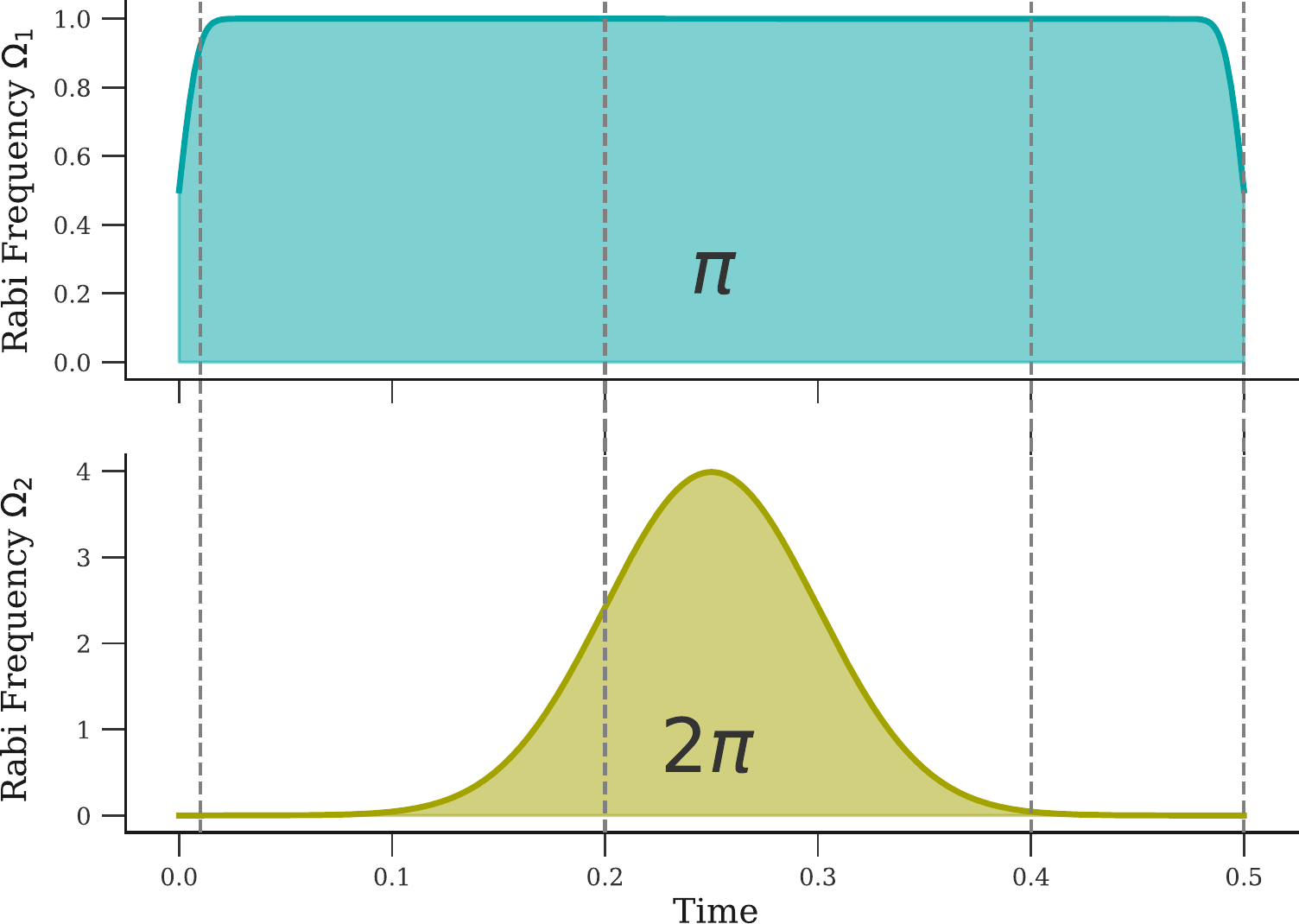}
    \caption{Pulse sequence for a simple simultaneous excitation scheme. A $\pi$ pulse is applied with Rabi frequency $\Omega_1$ to address the \transition{0}{1} transition (blue), and is applied weakly such that the pulse take the entire duration of the sequence to execute. Simultaneously to this, a stronger $2\pi$ pulse with Rabi frequency $\Omega_2$ addresses the \transition{1}{2} transition (gold) with a Gaussian intensity profile. The grey dashed lines mark the times with corresponding octant plots in figure \ref{fig:Qutrit Ramsey Result}.} \label{fig:Qutrit Ramsey Pulses}
\end{figure}

\begin{figure}[h]
    \centering
    \includegraphics[width = \linewidth]{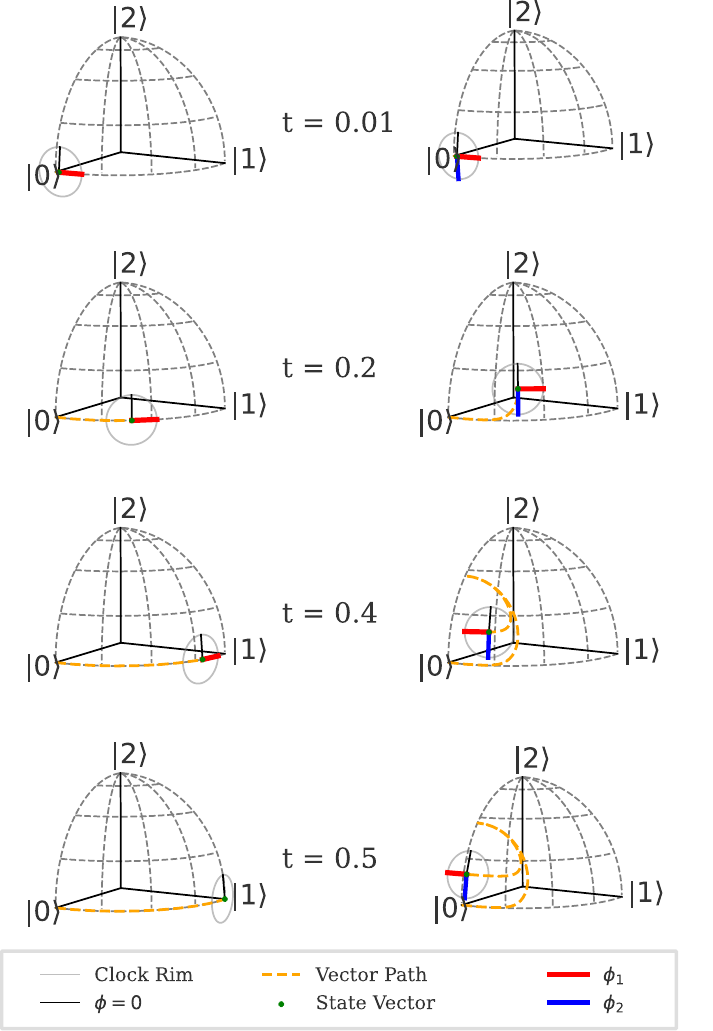}
    \caption{Octant plot for the sequence shown in figure \ref{fig:Qutrit Ramsey Pulses} for $\Omega_2 = 0$ (left column) and $\Omega_2 \neq 0$ (right column). The green line denoting the state population has been removed in favour of a point marker to reduce visual clutter in the diagram. \textbf{Left}: The sequence proceeds with a consistent transfer of population from \transition{0}{1}, ending with total transfer into $\ket{1}$. \textbf{Right}: The sequence proceeds with a transfer of population into $\ket{2}$, before partially de-exciting down to $\ket{1}$ and returning to no population in the intermediate $\ket{1}$ state by the end of the sequence.}
    \label{fig:Qutrit Ramsey Result}
\end{figure}

In order to show the effect of phase on a sequence, and how the octant plot can be illustrative in understanding it, a simple two pulse sequence was modelled using the Hamiltonian in (\ref{eq:Test Ham}) where the values of $\Omega_\mathrm{1}$ and $\Omega_\mathrm{2}$ are pulsed according to the sequence in figure \ref{fig:Qutrit Ramsey Pulses}.
This straightforward sequence consists of two pulses: a constant pulse addressing the \transition{0}{1} transition that acts as a $\pi$ pulse over the entire duration of the sequence, and a stronger pulse addressing the \transition{1}{2} transition with a Gaussian profile. Using these pulses, two cases were considered. The first was for the case where no $\Omega_2$ pulse is applied, and the qutrit is allowed to evolve stimulated only by the $\Omega_1$ pulse. The second is when this second $\Omega_2$ pulse is applied. The results for the cases of $\Omega_2 = 0$ and $\Omega_2 \neq 0$ are shown in the left and right hand columns of figure \ref{fig:Qutrit Ramsey Result} respectively.
\\
The sequence in the left column proceeds with a straightforward transfer of population from $\ket{0}$ to $\ket{1}$. As soon as the qutrit leaves the eigenstate $\ket{0}$ and the phasor $\phi_1$ becomes well defined it is immediately set to $\phi_1 = \frac{\pi}{2}$. This is  occurs because of the factor of $-i$ that is imprinted on the qutrit by the Schrödinger equation which dictates the time evolution of the system. As the Hamiltonian being applied is entirely real and with no diagonal terms, no further phase evolution occurs and $\phi_1$ remains unchanged for the duration of the sequence. This process is illustrative of the two cases where one would expect a reduced number of hands. For $t<0.5$, the state remains in a superposition between two states, leaving only one non-vanishing phasor. At $t = 0.5$ all population is in a single state, resulting in no non-vanishing phase terms. For the $\Omega_2 \neq 0$ case in the right hand column, the stimulation of the \transition{1}{2} transition causes all population to transfer out of $\ket{1}$ and into a superposition of $\ket{0}$ and $\ket{2}$. As $\phi_1$ is no longer defined at the point where this occurs, when population reenters $\ket{1}$ the phase is derived from $\phi_2$ with an additional $\frac{\pi}{2}$ radians again imprinted by the Schrödinger equation. Thus, the overall phase shift in $\phi_1$ is $\pi$ radians from the start of the sequence. This phase shift results in the qutrit evolving in antiphase with the qutrit in the $\Omega_2 = 0$ case. Thus, when it interacts with the driving field, it does so in opposition to the $\Omega_2 = 0$ qutrit and finishes the sequence with no population in $\ket{1}$. 
\\
\\
In visualising this sequence, the octant allows us to see the pivotal factor (the change in phase of $\phi_1$) that distinguishes the two cases, leaving one with total population transfer into $\ket{1}$ whilst leaving the other with no population in $\ket{1}$, despite both seeing the same $\Omega_1$ driving Rabi frequency throughout.
\section{Mixed State Description} \label{sec:Mixed States}
In an effectively noiseless, decay-free setting, the description in the previous section is adequate to fully describe any pure 3 level state. This, however, cannot generally be assumed. It is possible to engineer a system to be effectively pure by, for example, utilising either dressed \cite{Weidt2016Trapped-IonFields} or long-lived atomic states \cite{Maxwell2013StoragePolaritons}, but to create a visual description that fully encapsulates an arbitrary state we must look to extend the description in section \ref{sec:Pure States} to mixed states. 
\subsection{Theory} \label{subsec:Mixed Theory}
\begin{figure}[h]
    \centering
    \includegraphics[width = \linewidth]{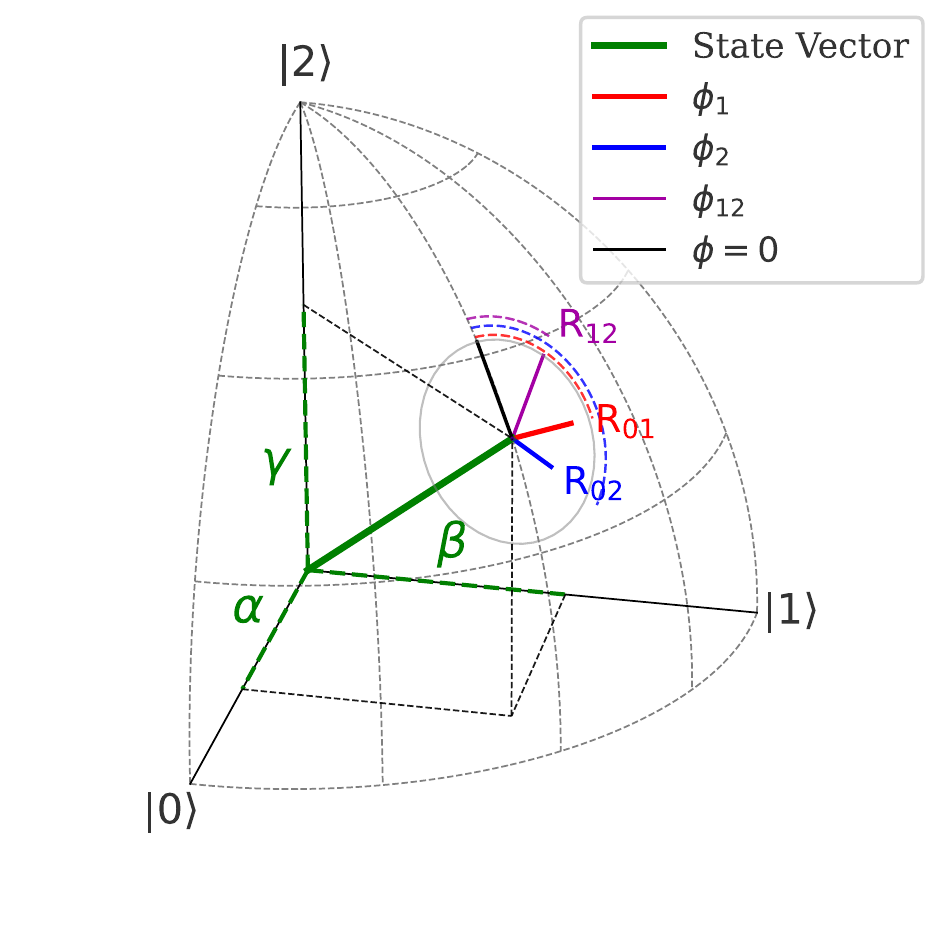}
    \caption{An example octant plot for the mixed state in (\ref{eq:Mixed Rho Example Case}). Here, $|\alpha| = |\beta| = |\gamma| = \frac{1}{\sqrt{3}}$, $\phi_{01} = \frac{\pi}{2}$, $\phi_{02} = \frac{3\pi}{4}$ and $\phi_{12} = \frac{6\pi}{5}$. The magnitudes of the clock hands are: $\mathrm{R}_{01} = \frac{3}{4}$,$\mathrm{R}_{02} = \frac{1}{2}$ and $\mathrm{R}_{12} = 1$. }
    \label{fig:Mixed State Example}
\end{figure}

\begin{figure}[h]
    \centering
    \includegraphics[width = 0.5\linewidth]{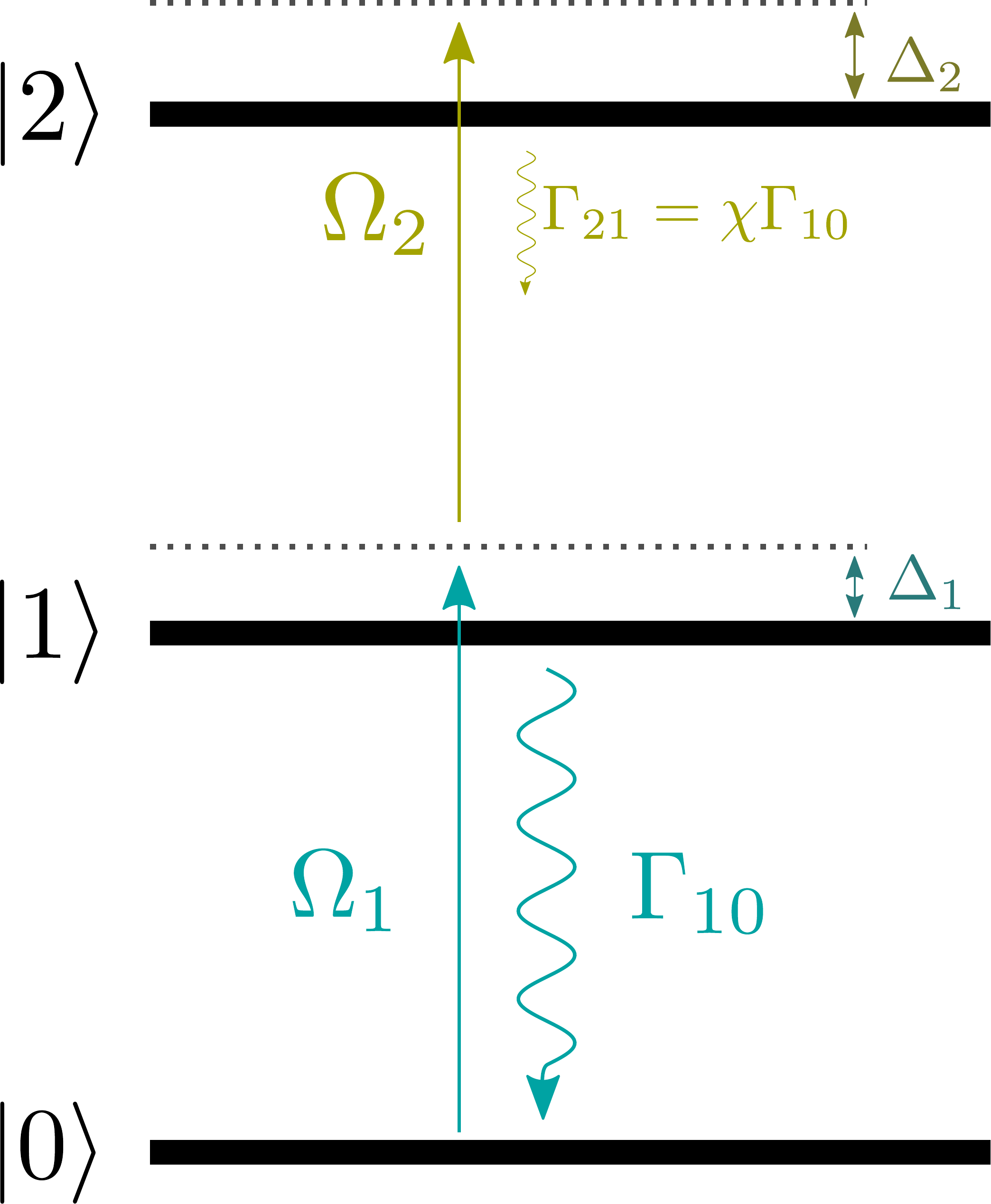}
    \caption{Atomic level scheme showing the variables, states and laser fields being considered in modelling. These are: the probe beam ($\Omega_1$) addressing the \transition{0}{1} transition, the coupling beam ($\Omega_2$) addressing the \transition{1}{2} transition, a pair of detuning terms ($\Delta_1$ and $\Delta_2$) and the decay rates $\Gamma_{10}$ and $\Gamma_{21}$. The coefficient $\chi$ is varied between simulations to adjust the extent of the loss of coherence in $\rho_{12} + \mathrm{c.c.}$.}
    \label{fig:Three Level Diagram}
\end{figure}

To extend to mixed states, we naturally have to adapt the description from pure state vectors to mixed state density matrices. The simplest of these is that of the pure state from (\ref{eq:Pure State})

\begin{equation} \label{eq:Pure Rho}
\rho = \ket{\psi} \! \bra{\psi} = \begin{pmatrix}
   |\alpha|^2 &\alpha \beta \mathrm{e}^{i\phi_1} &  \alpha \gamma \mathrm{e}^{i \phi_2}\\
    \alpha \beta \mathrm{e}^{-i\phi_1}& |\beta|^2 & \beta \gamma \mathrm{e}^{i(\phi_2-\phi_1)} \\
    \alpha \gamma \mathrm{e}^{-i \phi_2}& \beta \gamma \mathrm{e}^{-i(\phi_2-\phi_1)} & |\gamma|^2 \\
    \end{pmatrix}.
\end{equation}
To see where we might gain additional understanding over the pure state description, we compare this to a more arbitrary density matrix,
\begin{equation} \label{eq:Mixed Rho}
\rho = \begin{pmatrix}
   |\alpha|^2 & \mathrm{A} \mathrm{e}^{i\phi_{1}} &  \mathrm{B} \mathrm{e}^{i\phi_{2}}\\
    \mathrm{A} \mathrm{e}^{-i\phi_{1}} & |\beta|^2 & \mathrm{C} \mathrm{e}^{i\phi_{12}} \\
    \mathrm{B} \mathrm{e}^{i\phi_{2}} & \mathrm{C} \mathrm{e}^{-i\phi_{12}} & |\gamma|^2 \\
    \end{pmatrix}.
\end{equation}
Considering these matrices, we see that the phase term of the pure state coherence $\rho_{21}$ is only indicative of relative phase between $\ket{1}$ and $\ket{2}$ such that $\phi_{12}=  \phi_2-\phi_1$. In the cases of standard atomic radiative decay mechanisms and optical driving fields, which are the focus of the examples shown here, this relation holds in mixed states. Outside of such systems where extraneous decay mechanisms or 3 driving fields may be possible, this relation is not generally the case. Though such conditions are outside the scope of this work, displaying $\phi_{12}$ can reveal non-trivial information and warrants adding a third clock hand to the description. Additionally, assuming the presence of decay modes between states, the magnitude of the off-diagonal coherence terms ($\mathrm{A}, \mathrm{B}$ and $\mathrm{C}$) are also no longer trivial. To ensure that no information about the system is lost, the magnitudes of the coherence terms are encoded into the description as the length of the clock hands. In order to remain clear to the reader regardless of the absolute size of populations in the coherent states, these magnitudes ($\mathrm{R}_{jk}$) are normalised as
\begin{equation} \label{eq: Clock Hand Magnitude}
    \mathrm{R}_{jk}= \frac{|\rho_{jk}|}{\sqrt{|\rho_{jj}||\rho_{kk}|}}
\end{equation}
giving $0 \leq \mathrm{R}_{jk} \leq 1$. In the pure state case $|\rho_{01}|= \sqrt{\rho_{00}\rho_{11}} = \alpha \beta$ giving $\mathrm{R}_{01}$ and in a perfect statistical mixture $\rho_{\mathrm{mix}} = \mathrm{D}_0 \ket{0} \! \bra{0} + \mathrm{D}_1 \ket{1} \! \bra{1} + \mathrm{D}_2 \ket{2} \! \bra{2}$ this coherence term becomes $|\rho_{01}|= 0 \ \forall \ \mathrm{D}_j \in \mathbb{R}$. The same can also be applied to the other coherence terms $\rho_{j \neq k}$. This means that when a density matrix represents a complete statistical mixture as in $\rho_{\mathrm{mix}}$, this loss of coherence is represented not as a change in length of the octant plot's vector, but as a vanishing of the clock hands from the diagram. With the addition of the description of the state coherences, the diagram possesses the necessary 8 degrees of freedom (accounting for the trace condition $\rho_{00}+\rho_{11}+\rho_{22}=1$ removing a degree of freedom) required to fully express the $\mathrm{SU}(3)$ generator matrices that constitute any arbitrary qutrit state. A final adjustment to this description for cases where population decays out of the three levels (such as when coupled to a heat bath) would be to shorten the length of the state vector to account for the reduction of overall population in the system being considered. Examples of this effect will not be shown here and the discussion will be kept to decay between levels $\ket{0}$, $\ket{1}$ and $\ket{2}$. Thus this description allows a reader to easily interpret: the populations in each eigenstate, the relative phases between them and the degree of state purity via the size of the magnitudes $\mathrm{R}_{jk}$. An example for the mixed state 
\begin{equation} \label{eq:Mixed Rho Example Case}
\rho = \frac{1}{3}\begin{pmatrix}
   1 & \frac{3}{4} \mathrm{e}^{i\frac{\pi}{2}} &  \frac{1}{2} \mathrm{e}^{i\frac{3\pi}{4}}\\
    \frac{3}{4} \mathrm{e}^{-i\frac{\pi}{2}} & 1 & \mathrm{e}^{i\frac{\pi}{4}} \\
    \frac{1}{2} \mathrm{e}^{-i\frac{3\pi}{4}} & \mathrm{e}^{-i\frac{\pi}{4}} & 1 \\
    \end{pmatrix}.
\end{equation}
is shown in figure \ref{fig:Mixed State Example}.
\\
Though this visualisation method shows an individual qutrit state clearly, one shortcoming of the method is the lack of obvious distinction between orthogonal or distant states as measured by trace distance. Indeed, it is possible for the representations of two orthogonal states to share the same state vector on an octant plot; as is the case for $\ket{\psi}_+ = \frac{1}{\sqrt{2}} (\ket{0}+\ket{1})$ and $\ket{\psi}_- = \frac{1}{\sqrt{2}} (\ket{0}-\ket{1})$. Another shortcoming of this visualisation method is that by displaying the relative sizes of the coherent state magnitudes, we lack a measure of the absolute sizes of these off-diagonal terms. Thus the absolute sizes require deduction based on the position of the state vector of the octant and the sizes of the hands $\mathrm{R}_{jk}$. 
\\
For the remainder of this work, the three level system under consideration is that of a three level atom, and is shown in figure \ref{fig:Three Level Diagram}.
\\
In this system, state mixing is introduced via radiative decay between neighbouring states, causing a pure state density matrix to evolve into a statistical mixture over time. Mathematically, this is implemented by shifting from a time evolution of a state vector governed by the Schrödinger equation to a Lindbladian master equation of the form
\begin{equation} \label{eq:Lindblad Master}
\fulldiff{\rho}{t} = -i [\ham,\rho] + \sum_{j=1}^{2} \Gamma_{j,j-1} \left( \mathrm{C}_j \rho \mathrm{C}^\dag_j - \frac{1}{2} \{\rho,\mathrm{C}_j \mathrm{C}^\dag_j  \} \right)
\end{equation}
with collapse operators $\mathrm{C}_j = \ket{j-1} \bra{j}$ and $\hbar \equiv 1$. As in figure \ref{fig:Three Level Diagram} , the mixed state processes are performed with a strong decay mode between \transition{1}{0} with strength $\Gamma_{10}$, corresponding to decay modes present in atomic systems between neighbouring states. As this decay mode is the key defining feature of the system, values of other frequency parameters are quoted in terms of $\Gamma_{10}$, with times in $\tau_{10} = \frac{1}{\decay}$. the decay term $\Gamma_{21}$ is modelled to be a small fraction ($\chi$) of $\decay$ to simulate a long lived upper state such as a Rydberg state. In line with atomic physics, the Rabi frequencies $\Omega_1$ and $\Omega_2$ in figure \ref{fig:Three Level Diagram} correspond to probe and pump laser radiation fields addressing an atom respectively. By introducing a detuning of a laser field to the simulation, detuning terms ($\Delta_1$ and $\Delta_2$) can be introduced to the diagonals of the system Hamiltonian
\begin{equation} \label{eq:General Ham}
\ham = \begin{pmatrix}
   0 &\frac{\Omega_1}{2} & 0\\
    \frac{\Omega_1}{2}& -\Delta_1 & \frac{\Omega_2}{2} \\
    0& \frac{\Omega_2}{2} & -\Delta_{12} \\
    \end{pmatrix}
\end{equation}
where $\Delta_{12} = \Delta_1+\Delta_2$. The direct connection to atomic physics is made here such that the following examples of EIT and FWM can be accurately described in better context.
\subsection{Electromagnetically Induced Transparency} \label{subsec:EIT}
The EIT process is characterised by a sharp increase in the transmission of a weak probe beam through a medium under certain resonance conditions. This is present in the response of the medium as a reduced (or zero) excitation via the transition the probe stimulates. The Hamiltonian for this system is given in (\ref{eq:General Ham}).
 In the simplest case where $\Delta_1=\Delta_2 = 0$, the system forms an eigenstate, the so-called `dark state' only in terms of $\ket{0}$ and $\ket{2}$, and retains only a coherence in $\phi_{2}$. For driving fields $\Omega_1$ and $\Omega_2$ as shown in figure \ref{fig:Three Level Diagram} and (\ref{eq:General Ham}), the dark state $\ket{\psi}_{\mathrm{D}}$ is given by
\begin{equation} \label{eq:EIT Dark General}
\ket{\psi}_{\mathrm{D}} = \frac{1}{\sqrt{(\frac{\Omega_2}{\Omega_1})^2+1}} \begin{pmatrix}
    \frac{\Omega_2}{\Omega_1} \\
    0 \\
    -1
\end{pmatrix}.    
\end{equation}
 
\begin{figure}[h]
    \centering
    \includegraphics[width = \linewidth]{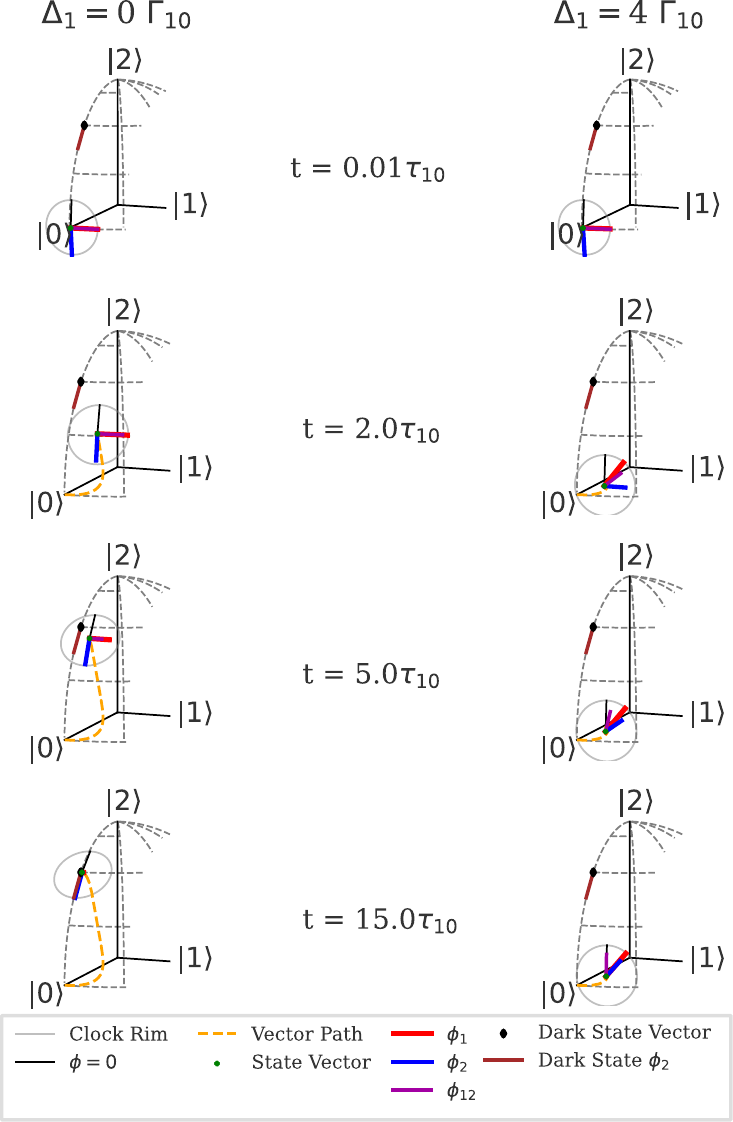}
    \caption{Time evolution of a qutrit state for resonant (left) and off-resonant (right) EIT. Here, $\Omega_1 = \Omega_2 = 2 \ \decay$ and $\Gamma_{21} = 1 \times 10^{-5} \ \decay$ which was chosen to make the \transition{2}{1} decay negligible on the timescale of the simulation. The solid black and pink lines shows the state vector and $\phi_{2}$ for the EIT dark state $\ket{\psi}_{\mathrm{D}}$ in (\ref{eq:EIT Dark Specific}).}
    \label{fig:EIT Detuning Mosaic}
\end{figure}

 This is in line with the values of $\rho_{01}$ and $\rho_{12}$ given in \cite{Gea-Banacloche1995ElectromagneticallyExperiment} for the case of zero detuning. When $\Delta_{1,2} \neq 0 $ we see that the eigenstates, though still analytically solvable, become much more complex and all contain a non-zero population in $\ket{1}$. The time evolution for a resonant (left column) and detuned (right column) EIT sequence are shown in figure \ref{fig:EIT Detuning Mosaic} for $\Omega_1=\Omega_2 = 2 \ \decay$. In these cases, a strong probe is modelled such that the dark state later described is visually distant from $\ket{0}$ for the sake of clear visualisation, but this is not typical in physical implementations. 
\\
\\
Similarly to the case in section \ref{subsec:Qutrit Ramsey}, the coherence states are initially populated with a $\frac{\pi}{2}$ phase shift imprinted by the Schrödinger equation. This results in the $\rho_{02}$ term immediately acquiring the necessary phase for the dark state rather than tending towards it over time. As the sequence progresses for the resonant case, an initial transient population transfer occurs into $\ket{1}$, with all three coherences present. As the sequence continues, this is transferred into $\ket{2}$ with any residual population in $\ket{1}$ decaying back to $\ket{0}$. The fact that this transfer back to $\ket{0}$ is a decay rather than coherent population transfer is shown by the loss of the $\phi_1$ clock hand before all population is fully transferred out of $\ket{1}$.  Alongside this decay of state population out of $\ket{1}$ and loss of $\rho_{10}$, the coherence term $\rho_{12}$ also decays to zero (shown by a vanishing $\phi_2$ hand) due to the loss of coherent population in $\ket{1}$, leaving only the $\rho_{02}$ coherence. This coherence remains at the value $\phi_{02} = \pi$, leaving the overall qutrit state in the EIT dark state. For the case of $\Omega_1 = \Omega_2 = 2 \ \decay$, the dark state (\ref{eq:EIT Dark General}) takes the specific form
\begin{equation} \label{eq:EIT Dark Specific}
\ket{\psi}_\mathrm{D} = \frac{1}{\sqrt{2}} (\ket{0} - \ket{2}).    
\end{equation}
This state is shown as the black diamond and brown clock hand in figure \ref{fig:EIT Detuning Mosaic}.
\\
For the off resonant case, the phases on the coherences $\phi_1$ and $\phi_2$ both tend towards $\phi_1= \phi_2 = \frac{\pi}{4}$, leaving the third phase difference term $\phi_{12}$ tending toward 0 throughout the simulation. As this becomes increasingly phase matched with the driving field, the response of the coherent state to the driving field decreases, preventing population from being driven into $\ket{2}$. Furthermore, in a similar vein to the resonant case, the coherences decrease over time albeit more gradually due to the comparatively small population being excited from $\ket{0}$. Using the octant plot, the dynamics of the EIT process can be clearly visualised. In particular, the coherences can be seen to decay without needing to interpret the small magnitudes of the coherence terms in a density matrix which, if not displayed analytically but instead with floating point variables (as is common for outputs in numerical simulations), may be hard to effectively and easily quantify at a glance. Not only is this magnitude easier to interpret than a numerically displayed complex number in any given off-diagonal, but the phase information is as well; the way that the $\phi_2$ phasor immediately becomes set to $\phi_2 = \pi$ once it is well defined, and that the phasor $\phi_{12} \rightarrow 0$ as $t \rightarrow \infty$ exemplify this in particular. 
\subsection{Four Wave Mixing}
\begin{figure}[h]
     \centering
     \includegraphics[width = \linewidth]{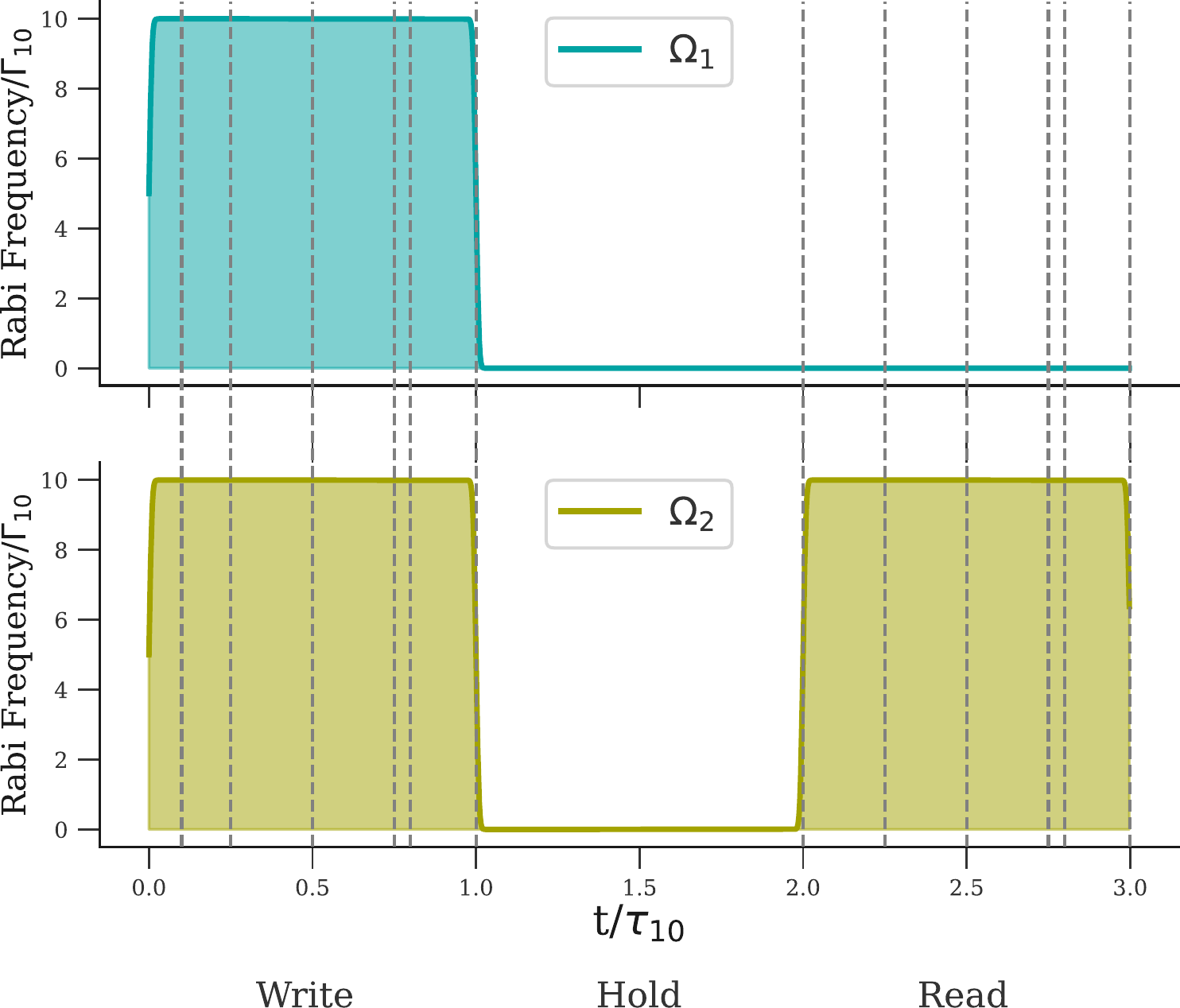}
     \caption{Rabi frequencies of the probe and coupling beams in a resonant FWM process. The grey dashed lines show the times during the write (read) stage that are shown in figure \ref{fig:FWM Write} (\ref{fig:FWM Read}).}
     \label{fig:FWM Pulses}
 \end{figure}

The final sequence of Four Wave Mixing (FWM) presented here again takes advantage of the decoherence effect of the decay modes in the system in figure \ref{fig:Three Level Diagram} to clearly display the rich dynamics at play. FWM is a process of great experimental relevance, with applications such as heralded single photon generation \cite{Lee2016HighlySystem} and coherent readout of stored photons \cite{Maxwell2013StoragePolaritons, Zugenmaier2018Long-livedTemperature}. FWM can in principle be performed with four states as in a diamond configuration as in \cite{Willis2009Four-waveVapor}, but 3 level ladder systems remain of significant interest \cite{Lee2017Single-photonEnsemble,Park2019Polarization-entangledInterferometer,Noh2021Four-waveStudy} and are thus worth discussion here. The pulse scheme considered here is to simulate the storage and retrieval of a photon in an atom and is split into three parts each of equal time $\mathrm{t} = \tau_{10}$. In the initial storage (write) time of $0 \leq t \leq \tau_{10}$, both driving fields are present, with the aim being to transfer population into the long-lived $\ket{2}$ state. In the second stage (hold) in the range $\tau_{10} < t < 2\tau_{10}$, no fields are present and the atom is free to decay via the decay modes $\decay$ and $\Gamma_{21}$. In the final stage (read) at times $2\tau_{10} < t < 3\tau_{10}$, driving by the Rabi frequency $\Omega_2$ is resumed, allowing the state to de-populate $\ket{2}$ and decay back to $\ket{0}$ via the strong $\decay$ decay mode. The values of $\Omega_1$ and $\Omega_2$ during this trio of write, hold and read steps are shown in figure \ref{fig:FWM Pulses}.
 
\begin{figure}[h]
     \centering
     \includegraphics[width = \linewidth]{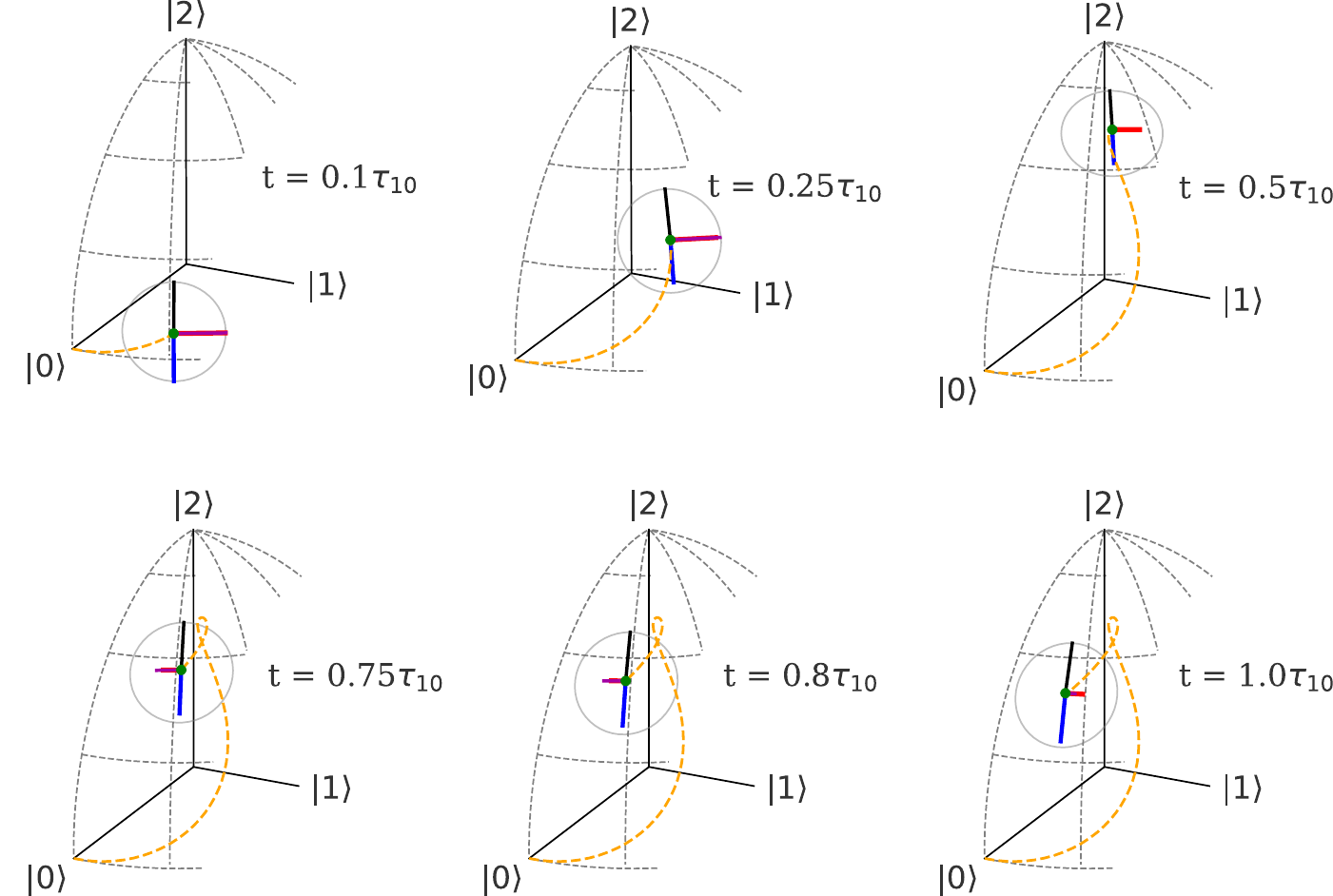}
     \caption{Initial dynamics of a FWM process with Rabi frequencies as indictated in the 'Write' stage of figure \ref{fig:FWM Pulses}. The system begins with an excitation of the qutrit to a long-lived upper state $\ket{2}$. The dotted orange line shows the path traced out by the state vector throughout the write sequence.}
     \label{fig:FWM Write}
 \end{figure}

\begin{figure}[h]
     \centering
     \includegraphics[width = \linewidth]{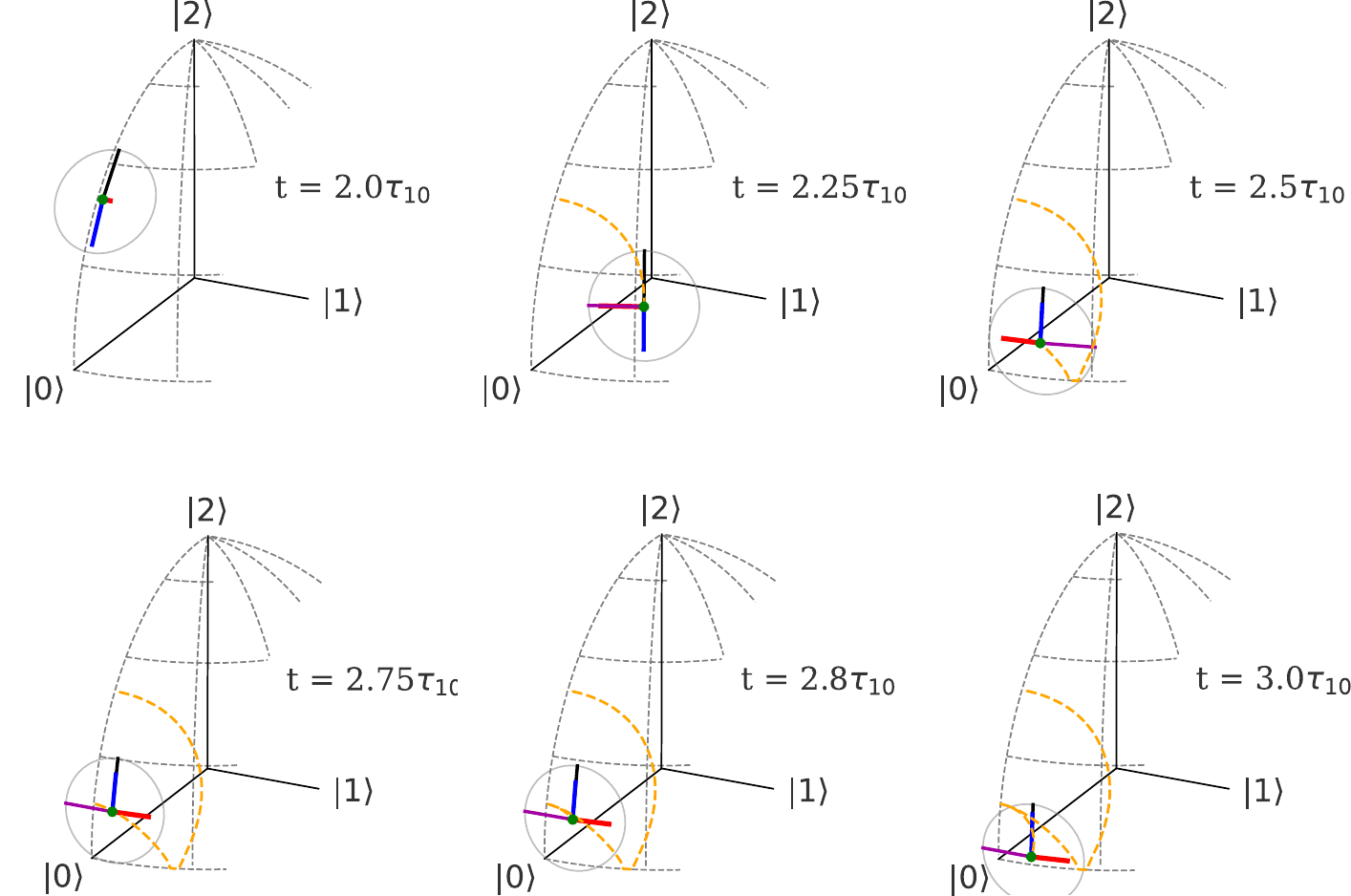}
     \caption{End of the FWM process described. Here, the qutrit de-excites from $\ket{2}$ state to the ground state via a strongly decaying intermediate state. The starting state shown here is different from the end state in figure \ref{fig:FWM Write} as an additional $\tau_{10}$ has passed between the two figures, causing the residual population in $\ket{1}$ to decay. The dotted orange line shows the path traced out by the state vector throughout the read sequence.}
     \label{fig:FWM Read}
 \end{figure}

 Throughout this sequence, the \transition{2}{1} decay mode is set to $\Gamma_{21} = 1 \times 10^{-3} \ \decay$ such that a slow decay occurs from \transition{2}{1} on the timescale of the simulation. The dashed grey lines in \ref{fig:FWM Pulses} during the write stage where both transitions are driven show the times for which octant plots are displayed in figure  \ref{fig:FWM Write}. For the second $\tau_{10}$, as neither transition is stimulated and the system is allowed to freely evolve, no plots are shown due to the trivial dynamics of simple \transition{1}{0} decay being the only dynamics of note. In the final $\tau_{10}$, where the $\Omega_{2}$ drives the upper \transition{1}{2} transition, the dashed grey lines show the times for which octant plots are rendered in figure \ref{fig:FWM Read}. As the coloured elements of the plot correspond to those in figure \ref{fig:EIT Detuning Mosaic} (excluding the dark state components), no additional legend is included in either of these plots.
 In the first two octants in figure \ref{fig:FWM Write}, an arc is swept out by the state vector due to the simultaneous driving of both transitions. An interesting feature present in these two, as well as the third plot, is the loss in coherence in the $\rho_{20}$ state as indicated by the shrinking blue hand despite the lack of a $\Gamma_{20}$. This feature can be accounted for by considering the $\rho_{20}$ term in (\ref{eq:Lindblad Master}) for no detunings and $\Omega_1 = \Omega_2 = 10 \decay$
 \begin{equation}
     \fulldiff{\rho_{20}}{t} = -i \cdot 10 \decay (\rho_{10} - \rho_{21}).
 \end{equation}
 Like the cases considered in both sections \ref{subsec:Qutrit Ramsey} and \ref{subsec:EIT}, we have that $\Re(\rho_{10})=\Re(\rho_{21}) = 0$. This results in a time evolution of the form 
 \begin{equation}
     \fulldiff{\rho_{20}}{t} = - 10 \decay (|\rho_{10}| - |\rho_{21}|)
 \end{equation}
 thus causing a decay in coherence dependent on the difference between the other coherence terms. Throughout this write sequence the coherence terms with decay modes also oscillate, albeit $\frac{\pi}{2}$ out of phase with the $\rho_{20}$ coherence and decaying in size throughout. In the $\tau_{10}$ time between figures \ref{fig:FWM Write} and \ref{fig:FWM Read} the only dynamic of note is the decay of population in the $\ket{1}$ state back to $\ket{0}$, resulting in the red $\phi_{1}$ hand vanishing entirely. The octant plots in figure \ref{fig:FWM Read} begin at the end ($t = 2 \tau_{10}$) of this hold stage.
 \\
During the final decay in figure \ref{fig:FWM Read}, population oscillates between the upper two states while decaying back to ground via state $\ket{1}$. Note that the $\phi_{1}$ hand returns when population is coherently recovered from the $\ket{2}$ between the first and second plots. The sudden changes in phase as the state oscillates towards $\ket{0}$ are again due to the depopulation of coherence terms resulting in only one phasor being well defined and the other phase terms thus acquiring a $\frac{\pi}{2}$ phase shift from this remaining coherence. The plots in the write sequence elucidate the dynamics of the $\rho_{20}$ term which may not initially be obvious when considering the level scheme shown in figure \ref{fig:Three Level Diagram}, and show that even though the other terms decay, the populations oscillate throughout the sequence. Again, these dynamics are not easy to intuit by considering the density matrix alone, and are only made clear by the octant plots for the process. For the latter read stage, the exact path that the qutrit takes back to ground state is shown alongside the phase jumps as the qutrit decays. Though these phase changes are not initially obvious, they have been discussed in the previous examples.
\section{Summary and Conclusions}
In this work, we have presented an intuitive formalism with which to visualise any arbitrary pure or mixed qutrit state for the purposes of education or for one to visually relay quantum dynamics in novel work. Test cases were examined to explore and illustrate their internal mechanics with non-trivial phase changes and coherence decays clearly illustrated by the octant plots. Though it is relatively straightforward to interpret the constituent elements of a 3 level density matrices, the aforementioned limitations of this visualisation method are not well explored here and could be presented in future work.  
\section{Acknowledgements}
Firstly, the author would like to thank Stuart Adams enormously for ideas of illustrative sequences and insights into the unfolding dynamics, many of which are shown here. Secondly, the advice from Nicholas Chancellor on key points worth highlighting and general guidance on narrative were enormously helpful in shaping this paper and also warrant a great deal of thanks. They gratefully thank Oliver Hughes, Karen Wadenpfuhl, Lucy Downes and Kevin Weatherill for countless conversations and pieces of useful feedback in preparation of this work. The author is also immensely grateful to Drs. Rodney and Frances Stubbs for the funding that made this work possible.  
\newpage
\bibliography{mendeleyreferences}

\begin{thebibliography}{40}%
\makeatletter
\providecommand \@ifxundefined [1]{%
 \@ifx{#1\undefined}
}%
\providecommand \@ifnum [1]{%
 \ifnum #1\expandafter \@firstoftwo
 \else \expandafter \@secondoftwo
 \fi
}%
\providecommand \@ifx [1]{%
 \ifx #1\expandafter \@firstoftwo
 \else \expandafter \@secondoftwo
 \fi
}%
\providecommand \natexlab [1]{#1}%
\providecommand \enquote  [1]{``#1''}%
\providecommand \bibnamefont  [1]{#1}%
\providecommand \bibfnamefont [1]{#1}%
\providecommand \citenamefont [1]{#1}%
\providecommand \href@noop [0]{\@secondoftwo}%
\providecommand \href [0]{\begingroup \@sanitize@url \@href}%
\providecommand \@href[1]{\@@startlink{#1}\@@href}%
\providecommand \@@href[1]{\endgroup#1\@@endlink}%
\providecommand \@sanitize@url [0]{\catcode `\\12\catcode `\$12\catcode
  `\&12\catcode `\#12\catcode `\^12\catcode `\_12\catcode `\%12\relax}%
\providecommand \@@startlink[1]{}%
\providecommand \@@endlink[0]{}%
\providecommand \url  [0]{\begingroup\@sanitize@url \@url }%
\providecommand \@url [1]{\endgroup\@href {#1}{\urlprefix }}%
\providecommand \urlprefix  [0]{URL }%
\providecommand \Eprint [0]{\href }%
\providecommand \doibase [0]{https://doi.org/}%
\providecommand \selectlanguage [0]{\@gobble}%
\providecommand \bibinfo  [0]{\@secondoftwo}%
\providecommand \bibfield  [0]{\@secondoftwo}%
\providecommand \translation [1]{[#1]}%
\providecommand \BibitemOpen [0]{}%
\providecommand \bibitemStop [0]{}%
\providecommand \bibitemNoStop [0]{.\EOS\space}%
\providecommand \EOS [0]{\spacefactor3000\relax}%
\providecommand \BibitemShut  [1]{\csname bibitem#1\endcsname}%
\let\auto@bib@innerbib\@empty
\bibitem [{\citenamefont {Shor}(1994)}]{Shor1994AlgorithmsFactoring}%
  \BibitemOpen
  \bibfield  {author} {\bibinfo {author} {\bibfnamefont {P.}~\bibnamefont
  {Shor}},\ }\bibfield  {title} {\bibinfo {title} {{Algorithms for quantum
  computation: discrete logarithms and factoring}},\ }in\ \href
  {https://doi.org/10.1109/SFCS.1994.365700} {\emph {\bibinfo {booktitle}
  {Proceedings 35th Annual Symposium on Foundations of Computer Science}}}\
  (\bibinfo  {publisher} {IEEE Comput. Soc. Press},\ \bibinfo {year} {1994})\
  pp.\ \bibinfo {pages} {124--134}\BibitemShut {NoStop}%
\bibitem [{\citenamefont {Grover}(1996)}]{Grover1996ASearch}%
  \BibitemOpen
  \bibfield  {author} {\bibinfo {author} {\bibfnamefont {L.~K.}\ \bibnamefont
  {Grover}},\ }\bibfield  {title} {\bibinfo {title} {{A fast quantum mechanical
  algorithm for database search}},\ }in\ \href
  {https://doi.org/10.1145/237814.237866} {\emph {\bibinfo {booktitle}
  {Proceedings of the twenty-eighth annual ACM symposium on Theory of computing
  - STOC '96}}}\ (\bibinfo  {publisher} {ACM Press},\ \bibinfo {address} {New
  York, New York, USA},\ \bibinfo {year} {1996})\ pp.\ \bibinfo {pages}
  {212--219}\BibitemShut {NoStop}%
\bibitem [{\citenamefont {Biamonte}\ \emph {et~al.}(2017)\citenamefont
  {Biamonte}, \citenamefont {Wittek}, \citenamefont {Pancotti}, \citenamefont
  {Rebentrost}, \citenamefont {Wiebe},\ and\ \citenamefont
  {Lloyd}}]{Biamonte2017QuantumLearning}%
  \BibitemOpen
  \bibfield  {author} {\bibinfo {author} {\bibfnamefont {J.}~\bibnamefont
  {Biamonte}}, \bibinfo {author} {\bibfnamefont {P.}~\bibnamefont {Wittek}},
  \bibinfo {author} {\bibfnamefont {N.}~\bibnamefont {Pancotti}}, \bibinfo
  {author} {\bibfnamefont {P.}~\bibnamefont {Rebentrost}}, \bibinfo {author}
  {\bibfnamefont {N.}~\bibnamefont {Wiebe}},\ and\ \bibinfo {author}
  {\bibfnamefont {S.}~\bibnamefont {Lloyd}},\ }\bibfield  {title} {\bibinfo
  {title} {{Quantum machine learning}},\ }\href
  {https://doi.org/10.1038/nature23474} {\bibfield  {journal} {\bibinfo
  {journal} {Nature}\ }\textbf {\bibinfo {volume} {549}},\ \bibinfo {pages}
  {195} (\bibinfo {year} {2017})}\BibitemShut {NoStop}%
\bibitem [{\citenamefont {Grigoryan}\ and\ \citenamefont
  {Pashayan}(2001)}]{Grigoryan2001AdiabaticDetuning}%
  \BibitemOpen
  \bibfield  {author} {\bibinfo {author} {\bibfnamefont {G.~G.}\ \bibnamefont
  {Grigoryan}}\ and\ \bibinfo {author} {\bibfnamefont {Y.~T.}\ \bibnamefont
  {Pashayan}},\ }\bibfield  {title} {\bibinfo {title} {{Adiabatic population
  transfer in three-level system with non-zero two-photon detuning}},\ }\href
  {https://doi.org/https://doi.org/10.1016/S0030-4018(01)01502-4} {\bibfield
  {journal} {\bibinfo  {journal} {Optics Communications}\ }\textbf {\bibinfo
  {volume} {198}},\ \bibinfo {pages} {107} (\bibinfo {year}
  {2001})}\BibitemShut {NoStop}%
\bibitem [{\citenamefont {Boradjiev}\ and\ \citenamefont
  {Vitanov}(2010)}]{Boradjiev2010StimulatedResonance}%
  \BibitemOpen
  \bibfield  {author} {\bibinfo {author} {\bibfnamefont {I.~I.}\ \bibnamefont
  {Boradjiev}}\ and\ \bibinfo {author} {\bibfnamefont {N.~V.}\ \bibnamefont
  {Vitanov}},\ }\bibfield  {title} {\bibinfo {title} {{Stimulated Raman
  adiabatic passage with unequal couplings: Beyond two-photon resonance}},\
  }\href {https://doi.org/10.1103/PhysRevA.81.053415} {\bibfield  {journal}
  {\bibinfo  {journal} {Physical Review A}\ }\textbf {\bibinfo {volume} {81}},\
  \bibinfo {pages} {053415} (\bibinfo {year} {2010})}\BibitemShut {NoStop}%
\bibitem [{\citenamefont {Miroshnychenko}\ \emph {et~al.}(2010)\citenamefont
  {Miroshnychenko}, \citenamefont {Ga{\"{e}}tan}, \citenamefont {Evellin},
  \citenamefont {Grangier}, \citenamefont {Comparat}, \citenamefont {Pillet},
  \citenamefont {Wilk},\ and\ \citenamefont
  {Browaeys}}]{Miroshnychenko2010CoherentState}%
  \BibitemOpen
  \bibfield  {author} {\bibinfo {author} {\bibfnamefont {Y.}~\bibnamefont
  {Miroshnychenko}}, \bibinfo {author} {\bibfnamefont {A.}~\bibnamefont
  {Ga{\"{e}}tan}}, \bibinfo {author} {\bibfnamefont {C.}~\bibnamefont
  {Evellin}}, \bibinfo {author} {\bibfnamefont {P.}~\bibnamefont {Grangier}},
  \bibinfo {author} {\bibfnamefont {D.}~\bibnamefont {Comparat}}, \bibinfo
  {author} {\bibfnamefont {P.}~\bibnamefont {Pillet}}, \bibinfo {author}
  {\bibfnamefont {T.}~\bibnamefont {Wilk}},\ and\ \bibinfo {author}
  {\bibfnamefont {A.}~\bibnamefont {Browaeys}},\ }\bibfield  {title} {\bibinfo
  {title} {{Coherent excitation of a single atom to a Rydberg state}},\ }\href
  {https://doi.org/10.1103/PhysRevA.82.013405} {\bibfield  {journal} {\bibinfo
  {journal} {Physical Review A}\ }\textbf {\bibinfo {volume} {82}},\ \bibinfo
  {pages} {013405} (\bibinfo {year} {2010})}\BibitemShut {NoStop}%
\bibitem [{\citenamefont {Cubel}\ \emph {et~al.}(2005)\citenamefont {Cubel},
  \citenamefont {Teo}, \citenamefont {Malinovsky}, \citenamefont {Guest},
  \citenamefont {Reinhard}, \citenamefont {Knuffman}, \citenamefont {Berman},\
  and\ \citenamefont {Raithel}}]{Cubel2005CoherentStates}%
  \BibitemOpen
  \bibfield  {author} {\bibinfo {author} {\bibfnamefont {T.}~\bibnamefont
  {Cubel}}, \bibinfo {author} {\bibfnamefont {B.~K.}\ \bibnamefont {Teo}},
  \bibinfo {author} {\bibfnamefont {V.~S.}\ \bibnamefont {Malinovsky}},
  \bibinfo {author} {\bibfnamefont {J.~R.}\ \bibnamefont {Guest}}, \bibinfo
  {author} {\bibfnamefont {A.}~\bibnamefont {Reinhard}}, \bibinfo {author}
  {\bibfnamefont {B.}~\bibnamefont {Knuffman}}, \bibinfo {author}
  {\bibfnamefont {P.~R.}\ \bibnamefont {Berman}},\ and\ \bibinfo {author}
  {\bibfnamefont {G.}~\bibnamefont {Raithel}},\ }\bibfield  {title} {\bibinfo
  {title} {{Coherent population transfer of ground-state atoms into Rydberg
  states}},\ }\href {https://doi.org/10.1103/PhysRevA.72.023405} {\bibfield
  {journal} {\bibinfo  {journal} {Physical Review A}\ }\textbf {\bibinfo
  {volume} {72}},\ \bibinfo {pages} {023405} (\bibinfo {year}
  {2005})}\BibitemShut {NoStop}%
\bibitem [{\citenamefont {Weidt}\ \emph {et~al.}(2016)\citenamefont {Weidt},
  \citenamefont {Randall}, \citenamefont {Webster}, \citenamefont {Lake},
  \citenamefont {Webb}, \citenamefont {Cohen}, \citenamefont {Navickas},
  \citenamefont {Lekitsch}, \citenamefont {Retzker},\ and\ \citenamefont
  {Hensinger}}]{Weidt2016Trapped-IonFields}%
  \BibitemOpen
  \bibfield  {author} {\bibinfo {author} {\bibfnamefont {S.}~\bibnamefont
  {Weidt}}, \bibinfo {author} {\bibfnamefont {J.}~\bibnamefont {Randall}},
  \bibinfo {author} {\bibfnamefont {S.}~\bibnamefont {Webster}}, \bibinfo
  {author} {\bibfnamefont {K.}~\bibnamefont {Lake}}, \bibinfo {author}
  {\bibfnamefont {A.}~\bibnamefont {Webb}}, \bibinfo {author} {\bibfnamefont
  {I.}~\bibnamefont {Cohen}}, \bibinfo {author} {\bibfnamefont
  {T.}~\bibnamefont {Navickas}}, \bibinfo {author} {\bibfnamefont
  {B.}~\bibnamefont {Lekitsch}}, \bibinfo {author} {\bibfnamefont
  {A.}~\bibnamefont {Retzker}},\ and\ \bibinfo {author} {\bibfnamefont
  {W.}~\bibnamefont {Hensinger}},\ }\bibfield  {title} {\bibinfo {title}
  {{Trapped-Ion Quantum Logic with Global Radiation Fields}},\ }\href
  {https://doi.org/10.1103/PhysRevLett.117.220501} {\bibfield  {journal}
  {\bibinfo  {journal} {Physical Review Letters}\ }\textbf {\bibinfo {volume}
  {117}},\ \bibinfo {pages} {220501} (\bibinfo {year} {2016})}\BibitemShut
  {NoStop}%
\bibitem [{\citenamefont {Molony}\ \emph {et~al.}(2014)\citenamefont {Molony},
  \citenamefont {Gregory}, \citenamefont {Ji}, \citenamefont {Lu},
  \citenamefont {K{\"{o}}ppinger}, \citenamefont {Le~Sueur}, \citenamefont
  {Blackley}, \citenamefont {Hutson},\ and\ \citenamefont
  {Cornish}}]{Molony2014CreationState}%
  \BibitemOpen
  \bibfield  {author} {\bibinfo {author} {\bibfnamefont {P.~K.}\ \bibnamefont
  {Molony}}, \bibinfo {author} {\bibfnamefont {P.~D.}\ \bibnamefont {Gregory}},
  \bibinfo {author} {\bibfnamefont {Z.}~\bibnamefont {Ji}}, \bibinfo {author}
  {\bibfnamefont {B.}~\bibnamefont {Lu}}, \bibinfo {author} {\bibfnamefont
  {M.~P.}\ \bibnamefont {K{\"{o}}ppinger}}, \bibinfo {author} {\bibfnamefont
  {C.~R.}\ \bibnamefont {Le~Sueur}}, \bibinfo {author} {\bibfnamefont {C.~L.}\
  \bibnamefont {Blackley}}, \bibinfo {author} {\bibfnamefont {J.~M.}\
  \bibnamefont {Hutson}},\ and\ \bibinfo {author} {\bibfnamefont {S.~L.}\
  \bibnamefont {Cornish}},\ }\bibfield  {title} {\bibinfo {title} {{Creation of
  Ultracold 87Rb 133Cs Molecules in the Rovibrational Ground State}},\ }\href
  {https://doi.org/10.1103/PhysRevLett.113.255301} {\bibfield  {journal}
  {\bibinfo  {journal} {Physical Review Letters}\ }\textbf {\bibinfo {volume}
  {113}},\ \bibinfo {pages} {255301} (\bibinfo {year} {2014})}\BibitemShut
  {NoStop}%
\bibitem [{\citenamefont {Menchon-Enrich}\ \emph {et~al.}(2016)\citenamefont
  {Menchon-Enrich}, \citenamefont {Benseny}, \citenamefont {Ahufinger},
  \citenamefont {Greentree}, \citenamefont {Busch},\ and\ \citenamefont
  {Mompart}}]{Menchon-Enrich2016SpatialProgress}%
  \BibitemOpen
  \bibfield  {author} {\bibinfo {author} {\bibfnamefont {R.}~\bibnamefont
  {Menchon-Enrich}}, \bibinfo {author} {\bibfnamefont {A.}~\bibnamefont
  {Benseny}}, \bibinfo {author} {\bibfnamefont {V.}~\bibnamefont {Ahufinger}},
  \bibinfo {author} {\bibfnamefont {A.~D.}\ \bibnamefont {Greentree}}, \bibinfo
  {author} {\bibfnamefont {T.}~\bibnamefont {Busch}},\ and\ \bibinfo {author}
  {\bibfnamefont {J.}~\bibnamefont {Mompart}},\ }\bibfield  {title} {\bibinfo
  {title} {{Spatial adiabatic passage: a review of recent progress}},\ }\href
  {https://doi.org/10.1088/0034-4885/79/7/074401} {\bibfield  {journal}
  {\bibinfo  {journal} {Reports on Progress in Physics}\ }\textbf {\bibinfo
  {volume} {79}},\ \bibinfo {pages} {074401} (\bibinfo {year}
  {2016})}\BibitemShut {NoStop}%
\bibitem [{\citenamefont {Fleischhauer}\ \emph {et~al.}(2005)\citenamefont
  {Fleischhauer}, \citenamefont {Imamoglu},\ and\ \citenamefont
  {Marangos}}]{Fleischhauer2005ElectromagneticallyMedia}%
  \BibitemOpen
  \bibfield  {author} {\bibinfo {author} {\bibfnamefont {M.}~\bibnamefont
  {Fleischhauer}}, \bibinfo {author} {\bibfnamefont {A.}~\bibnamefont
  {Imamoglu}},\ and\ \bibinfo {author} {\bibfnamefont {J.~P.}\ \bibnamefont
  {Marangos}},\ }\bibfield  {title} {\bibinfo {title} {{Electromagnetically
  induced transparency: Optics in coherent media}},\ }\href
  {https://doi.org/https://doi.org/10.1103/RevModPhys.77.633} {\bibfield
  {journal} {\bibinfo  {journal} {Reviews of Modern Physics}\ }\textbf
  {\bibinfo {volume} {77}},\ \bibinfo {pages} {633} (\bibinfo {year}
  {2005})}\BibitemShut {NoStop}%
\bibitem [{\citenamefont {Schraft}\ \emph {et~al.}(2016)\citenamefont
  {Schraft}, \citenamefont {Hain}, \citenamefont {Lorenz},\ and\ \citenamefont
  {Halfmann}}]{Schraft2016StoppedCrystal}%
  \BibitemOpen
  \bibfield  {author} {\bibinfo {author} {\bibfnamefont {D.}~\bibnamefont
  {Schraft}}, \bibinfo {author} {\bibfnamefont {M.}~\bibnamefont {Hain}},
  \bibinfo {author} {\bibfnamefont {N.}~\bibnamefont {Lorenz}},\ and\ \bibinfo
  {author} {\bibfnamefont {T.}~\bibnamefont {Halfmann}},\ }\bibfield  {title}
  {\bibinfo {title} {{Stopped Light at High Storage Efficiency in a Pr3+:Y2SiO5
  Crystal}},\ }\href {https://doi.org/10.1103/PhysRevLett.116.073602}
  {\bibfield  {journal} {\bibinfo  {journal} {Physical Review Letters}\
  }\textbf {\bibinfo {volume} {116}},\ \bibinfo {pages} {073602} (\bibinfo
  {year} {2016})}\BibitemShut {NoStop}%
\bibitem [{\citenamefont {Nicolas}\ \emph {et~al.}(2014)\citenamefont
  {Nicolas}, \citenamefont {Veissier}, \citenamefont {Giner}, \citenamefont
  {Giacobino}, \citenamefont {Maxein},\ and\ \citenamefont
  {Laurat}}]{Nicolas2014AQubits}%
  \BibitemOpen
  \bibfield  {author} {\bibinfo {author} {\bibfnamefont {A.}~\bibnamefont
  {Nicolas}}, \bibinfo {author} {\bibfnamefont {L.}~\bibnamefont {Veissier}},
  \bibinfo {author} {\bibfnamefont {L.}~\bibnamefont {Giner}}, \bibinfo
  {author} {\bibfnamefont {E.}~\bibnamefont {Giacobino}}, \bibinfo {author}
  {\bibfnamefont {D.}~\bibnamefont {Maxein}},\ and\ \bibinfo {author}
  {\bibfnamefont {J.}~\bibnamefont {Laurat}},\ }\bibfield  {title} {\bibinfo
  {title} {{A quantum memory for orbital angular momentum photonic qubits}},\
  }\href {https://doi.org/10.1038/nphoton.2013.355} {\bibfield  {journal}
  {\bibinfo  {journal} {Nature Photonics}\ }\textbf {\bibinfo {volume} {8}},\
  \bibinfo {pages} {234} (\bibinfo {year} {2014})}\BibitemShut {NoStop}%
\bibitem [{\citenamefont {Lin}\ \emph {et~al.}(2016)\citenamefont {Lin},
  \citenamefont {Chen}, \citenamefont {Yu}, \citenamefont {Liu}, \citenamefont
  {Li},\ and\ \citenamefont {Chen}}]{Lin2016AnFunctions}%
  \BibitemOpen
  \bibfield  {author} {\bibinfo {author} {\bibfnamefont {X.~Q.}\ \bibnamefont
  {Lin}}, \bibinfo {author} {\bibfnamefont {Z.}~\bibnamefont {Chen}}, \bibinfo
  {author} {\bibfnamefont {J.~W.}\ \bibnamefont {Yu}}, \bibinfo {author}
  {\bibfnamefont {P.~Q.}\ \bibnamefont {Liu}}, \bibinfo {author} {\bibfnamefont
  {P.~F.}\ \bibnamefont {Li}},\ and\ \bibinfo {author} {\bibfnamefont
  {Z.}~\bibnamefont {Chen}},\ }\bibfield  {title} {\bibinfo {title} {{An
  EIT-Based Compact Microwave Sensor With Double Sensing Functions}},\ }\href
  {https://doi.org/10.1109/JSEN.2015.2480800} {\bibfield  {journal} {\bibinfo
  {journal} {IEEE Sensors Journal}\ }\textbf {\bibinfo {volume} {16}},\
  \bibinfo {pages} {293} (\bibinfo {year} {2016})}\BibitemShut {NoStop}%
\bibitem [{\citenamefont {Kleinman}(1962)}]{Kleinman1962TheoryLight}%
  \BibitemOpen
  \bibfield  {author} {\bibinfo {author} {\bibfnamefont {D.~A.}\ \bibnamefont
  {Kleinman}},\ }\bibfield  {title} {\bibinfo {title} {{Theory of Second
  Harmonic Generation of Light}},\ }\href
  {https://doi.org/10.1103/PhysRev.128.1761} {\bibfield  {journal} {\bibinfo
  {journal} {Physical Review}\ }\textbf {\bibinfo {volume} {128}},\ \bibinfo
  {pages} {1761} (\bibinfo {year} {1962})}\BibitemShut {NoStop}%
\bibitem [{\citenamefont {Lin}\ \emph {et~al.}(1999)\citenamefont {Lin},
  \citenamefont {Lee}, \citenamefont {Liu}, \citenamefont {Chen},\ and\
  \citenamefont {Pickard}}]{Lin1999MechanismCrystals}%
  \BibitemOpen
  \bibfield  {author} {\bibinfo {author} {\bibfnamefont {J.}~\bibnamefont
  {Lin}}, \bibinfo {author} {\bibfnamefont {M.-H.}\ \bibnamefont {Lee}},
  \bibinfo {author} {\bibfnamefont {Z.-P.}\ \bibnamefont {Liu}}, \bibinfo
  {author} {\bibfnamefont {C.}~\bibnamefont {Chen}},\ and\ \bibinfo {author}
  {\bibfnamefont {C.~J.}\ \bibnamefont {Pickard}},\ }\bibfield  {title}
  {\bibinfo {title} {{Mechanism for linear and nonlinear optical effects in
  {$\beta$}−BaB2O4 crystals}},\ }\href
  {https://doi.org/10.1103/PhysRevB.60.13380} {\bibfield  {journal} {\bibinfo
  {journal} {Physical Review B}\ }\textbf {\bibinfo {volume} {60}},\ \bibinfo
  {pages} {13380} (\bibinfo {year} {1999})}\BibitemShut {NoStop}%
\bibitem [{\citenamefont {Cong}\ \emph {et~al.}(2012)\citenamefont {Cong},
  \citenamefont {Yang}, \citenamefont {Liao}, \citenamefont {Wang},
  \citenamefont {Lin},\ and\ \citenamefont {Lin}}]{Cong2012Experimental3OH}%
  \BibitemOpen
  \bibfield  {author} {\bibinfo {author} {\bibfnamefont {R.}~\bibnamefont
  {Cong}}, \bibinfo {author} {\bibfnamefont {T.}~\bibnamefont {Yang}}, \bibinfo
  {author} {\bibfnamefont {F.}~\bibnamefont {Liao}}, \bibinfo {author}
  {\bibfnamefont {Y.}~\bibnamefont {Wang}}, \bibinfo {author} {\bibfnamefont
  {Z.}~\bibnamefont {Lin}},\ and\ \bibinfo {author} {\bibfnamefont
  {J.}~\bibnamefont {Lin}},\ }\bibfield  {title} {\bibinfo {title}
  {{Experimental and theoretical studies of second harmonic generation for Bi
  2O 2[NO 3(OH)]}},\ }\href
  {https://doi.org/10.1016/j.materresbull.2012.04.145} {\bibfield  {journal}
  {\bibinfo  {journal} {Materials Research Bulletin}\ }\textbf {\bibinfo
  {volume} {47}},\ \bibinfo {pages} {2573} (\bibinfo {year}
  {2012})}\BibitemShut {NoStop}%
\bibitem [{\citenamefont {Lee}\ \emph {et~al.}(2016)\citenamefont {Lee},
  \citenamefont {Lee}, \citenamefont {Kim},\ and\ \citenamefont
  {Moon}}]{Lee2016HighlySystem}%
  \BibitemOpen
  \bibfield  {author} {\bibinfo {author} {\bibfnamefont {Y.-S.}\ \bibnamefont
  {Lee}}, \bibinfo {author} {\bibfnamefont {S.~M.}\ \bibnamefont {Lee}},
  \bibinfo {author} {\bibfnamefont {H.}~\bibnamefont {Kim}},\ and\ \bibinfo
  {author} {\bibfnamefont {H.~S.}\ \bibnamefont {Moon}},\ }\bibfield  {title}
  {\bibinfo {title} {{Highly bright photon-pair generation in Doppler-broadened
  ladder-type atomic system}},\ }\href {https://doi.org/10.1364/oe.24.028083}
  {\bibfield  {journal} {\bibinfo  {journal} {Optics Express}\ }\textbf
  {\bibinfo {volume} {24}},\ \bibinfo {pages} {28083} (\bibinfo {year}
  {2016})}\BibitemShut {NoStop}%
\bibitem [{\citenamefont {Maxwell}\ \emph {et~al.}(2013)\citenamefont
  {Maxwell}, \citenamefont {Szwer}, \citenamefont {Paredes-Barato},
  \citenamefont {Busche}, \citenamefont {Pritchard}, \citenamefont {Gauguet},
  \citenamefont {Weatherill}, \citenamefont {Jones},\ and\ \citenamefont
  {Adams}}]{Maxwell2013StoragePolaritons}%
  \BibitemOpen
  \bibfield  {author} {\bibinfo {author} {\bibfnamefont {D.}~\bibnamefont
  {Maxwell}}, \bibinfo {author} {\bibfnamefont {D.~J.}\ \bibnamefont {Szwer}},
  \bibinfo {author} {\bibfnamefont {D.}~\bibnamefont {Paredes-Barato}},
  \bibinfo {author} {\bibfnamefont {H.}~\bibnamefont {Busche}}, \bibinfo
  {author} {\bibfnamefont {J.~D.}\ \bibnamefont {Pritchard}}, \bibinfo {author}
  {\bibfnamefont {A.}~\bibnamefont {Gauguet}}, \bibinfo {author} {\bibfnamefont
  {K.~J.}\ \bibnamefont {Weatherill}}, \bibinfo {author} {\bibfnamefont
  {M.~P.~A.}\ \bibnamefont {Jones}},\ and\ \bibinfo {author} {\bibfnamefont
  {C.~S.}\ \bibnamefont {Adams}},\ }\bibfield  {title} {\bibinfo {title}
  {{Storage and Control of Optical Photons Using Rydberg Polaritons}},\ }\href
  {https://doi.org/10.1103/PhysRevLett.110.103001} {\bibfield  {journal}
  {\bibinfo  {journal} {Physical Review Letters}\ }\textbf {\bibinfo {volume}
  {110}},\ \bibinfo {pages} {103001} (\bibinfo {year} {2013})}\BibitemShut
  {NoStop}%
\bibitem [{\citenamefont {Zugenmaier}\ \emph {et~al.}(2018)\citenamefont
  {Zugenmaier}, \citenamefont {Dideriksen}, \citenamefont {S{\o}rensen},
  \citenamefont {Albrecht},\ and\ \citenamefont
  {Polzik}}]{Zugenmaier2018Long-livedTemperature}%
  \BibitemOpen
  \bibfield  {author} {\bibinfo {author} {\bibfnamefont {M.}~\bibnamefont
  {Zugenmaier}}, \bibinfo {author} {\bibfnamefont {K.~B.}\ \bibnamefont
  {Dideriksen}}, \bibinfo {author} {\bibfnamefont {A.~S.}\ \bibnamefont
  {S{\o}rensen}}, \bibinfo {author} {\bibfnamefont {B.}~\bibnamefont
  {Albrecht}},\ and\ \bibinfo {author} {\bibfnamefont {E.~S.}\ \bibnamefont
  {Polzik}},\ }\bibfield  {title} {\bibinfo {title} {{Long-lived non-classical
  correlations towards quantum communication at room temperature}},\ }\href
  {https://doi.org/10.1038/s42005-018-0080-x} {\bibfield  {journal} {\bibinfo
  {journal} {Communications Physics}\ }\textbf {\bibinfo {volume} {1}},\
  \bibinfo {pages} {76} (\bibinfo {year} {2018})}\BibitemShut {NoStop}%
\bibitem [{\citenamefont {Willis}\ \emph {et~al.}(2009)\citenamefont {Willis},
  \citenamefont {Becerra}, \citenamefont {Orozco},\ and\ \citenamefont
  {Rolston}}]{Willis2009Four-waveVapor}%
  \BibitemOpen
  \bibfield  {author} {\bibinfo {author} {\bibfnamefont {R.~T.}\ \bibnamefont
  {Willis}}, \bibinfo {author} {\bibfnamefont {F.~E.}\ \bibnamefont {Becerra}},
  \bibinfo {author} {\bibfnamefont {L.~A.}\ \bibnamefont {Orozco}},\ and\
  \bibinfo {author} {\bibfnamefont {S.~L.}\ \bibnamefont {Rolston}},\
  }\bibfield  {title} {\bibinfo {title} {{Four-wave mixing in the diamond
  configuration in an atomic vapor}},\ }\href
  {https://doi.org/10.1103/PhysRevA.79.033814} {\bibfield  {journal} {\bibinfo
  {journal} {Physical Review A}\ }\textbf {\bibinfo {volume} {79}},\ \bibinfo
  {pages} {033814} (\bibinfo {year} {2009})}\BibitemShut {NoStop}%
\bibitem [{\citenamefont {Lee}\ \emph {et~al.}(2017)\citenamefont {Lee},
  \citenamefont {Lee}, \citenamefont {Kim},\ and\ \citenamefont
  {Moon}}]{Lee2017Single-photonEnsemble}%
  \BibitemOpen
  \bibfield  {author} {\bibinfo {author} {\bibfnamefont {Y.-S.}\ \bibnamefont
  {Lee}}, \bibinfo {author} {\bibfnamefont {S.~M.}\ \bibnamefont {Lee}},
  \bibinfo {author} {\bibfnamefont {H.}~\bibnamefont {Kim}},\ and\ \bibinfo
  {author} {\bibfnamefont {H.~S.}\ \bibnamefont {Moon}},\ }\bibfield  {title}
  {\bibinfo {title} {{Single-photon superradiant beating from a
  Doppler-broadened ladder-type atomic ensemble}},\ }\href
  {https://doi.org/10.1103/PhysRevA.96.063832} {\bibfield  {journal} {\bibinfo
  {journal} {Physical Review A}\ }\textbf {\bibinfo {volume} {96}},\ \bibinfo
  {pages} {063832} (\bibinfo {year} {2017})}\BibitemShut {NoStop}%
\bibitem [{\citenamefont {Park}\ \emph {et~al.}(2019)\citenamefont {Park},
  \citenamefont {Kim},\ and\ \citenamefont
  {Moon}}]{Park2019Polarization-entangledInterferometer}%
  \BibitemOpen
  \bibfield  {author} {\bibinfo {author} {\bibfnamefont {J.}~\bibnamefont
  {Park}}, \bibinfo {author} {\bibfnamefont {H.}~\bibnamefont {Kim}},\ and\
  \bibinfo {author} {\bibfnamefont {H.~S.}\ \bibnamefont {Moon}},\ }\bibfield
  {title} {\bibinfo {title} {{Polarization-Entangled Photons from a Warm Atomic
  Ensemble Using a Sagnac Interferometer}},\ }\href
  {https://doi.org/10.1103/PhysRevLett.122.143601} {\bibfield  {journal}
  {\bibinfo  {journal} {Physical Review Letters}\ }\textbf {\bibinfo {volume}
  {122}},\ \bibinfo {pages} {143601} (\bibinfo {year} {2019})}\BibitemShut
  {NoStop}%
\bibitem [{\citenamefont {Noh}\ and\ \citenamefont
  {Seb~Moon}(2021)}]{Noh2021Four-waveStudy}%
  \BibitemOpen
  \bibfield  {author} {\bibinfo {author} {\bibfnamefont {H.-R.}\ \bibnamefont
  {Noh}}\ and\ \bibinfo {author} {\bibfnamefont {H.}~\bibnamefont {Seb~Moon}},\
  }\bibfield  {title} {\bibinfo {title} {{Four-wave mixing in a ladder
  configuration of warm 87Rb atoms: a theoretical study}},\ }\href
  {https://doi.org/10.1364/OE.416960} {\bibfield  {journal} {\bibinfo
  {journal} {Optics Express}\ }\textbf {\bibinfo {volume} {29}},\ \bibinfo
  {pages} {6495} (\bibinfo {year} {2021})}\BibitemShut {NoStop}%
\bibitem [{\citenamefont {Nielsen}\ and\ \citenamefont
  {Chuang}(2010)}]{Nielsen2010QuantumInformation}%
  \BibitemOpen
  \bibfield  {author} {\bibinfo {author} {\bibfnamefont {M.~A.}\ \bibnamefont
  {Nielsen}}\ and\ \bibinfo {author} {\bibfnamefont {I.~L.}\ \bibnamefont
  {Chuang}},\ }\href@noop {} {\emph {\bibinfo {title} {{Quantum computation and
  quantum information}}}}\ (\bibinfo  {publisher} {Cambridge University
  Press},\ \bibinfo {year} {2010})\BibitemShut {NoStop}%
\bibitem [{\citenamefont {Coecke}\ and\ \citenamefont
  {Duncan}(2011)}]{Coecke2011InteractingDiagrammatics}%
  \BibitemOpen
  \bibfield  {author} {\bibinfo {author} {\bibfnamefont {B.}~\bibnamefont
  {Coecke}}\ and\ \bibinfo {author} {\bibfnamefont {R.}~\bibnamefont
  {Duncan}},\ }\bibfield  {title} {\bibinfo {title} {{Interacting quantum
  observables: categorical algebra and diagrammatics}},\ }\href
  {https://doi.org/10.1088/1367-2630/13/4/043016} {\bibfield  {journal}
  {\bibinfo  {journal} {New Journal of Physics}\ }\textbf {\bibinfo {volume}
  {13}},\ \bibinfo {pages} {043016} (\bibinfo {year} {2011})}\BibitemShut
  {NoStop}%
\bibitem [{\citenamefont {Kissinger}\ and\ \citenamefont {van~de
  Wetering}(2019)}]{Kissinger2019UniversalMeasurements}%
  \BibitemOpen
  \bibfield  {author} {\bibinfo {author} {\bibfnamefont {A.}~\bibnamefont
  {Kissinger}}\ and\ \bibinfo {author} {\bibfnamefont {J.}~\bibnamefont {van~de
  Wetering}},\ }\bibfield  {title} {\bibinfo {title} {{Universal MBQC with
  generalised parity-phase interactions and Pauli measurements}},\ }\href
  {https://doi.org/10.22331/q-2019-04-26-134} {\bibfield  {journal} {\bibinfo
  {journal} {Quantum}\ }\textbf {\bibinfo {volume} {3}},\ \bibinfo {pages}
  {134} (\bibinfo {year} {2019})}\BibitemShut {NoStop}%
\bibitem [{\citenamefont {de~Beaudrap}\ and\ \citenamefont
  {Horsman}(2020)}]{deBeaudrap2020TheSurgery}%
  \BibitemOpen
  \bibfield  {author} {\bibinfo {author} {\bibfnamefont {N.}~\bibnamefont
  {de~Beaudrap}}\ and\ \bibinfo {author} {\bibfnamefont {D.}~\bibnamefont
  {Horsman}},\ }\bibfield  {title} {\bibinfo {title} {{The ZX calculus is a
  language for surface code lattice surgery}},\ }\href
  {https://doi.org/10.22331/q-2020-01-09-218} {\bibfield  {journal} {\bibinfo
  {journal} {Quantum}\ }\textbf {\bibinfo {volume} {4}},\ \bibinfo {pages}
  {218} (\bibinfo {year} {2020})}\BibitemShut {NoStop}%
\bibitem [{\citenamefont {Kissinger}\ and\ \citenamefont {van~de
  Wetering}(2022)}]{Kissinger2022SimulatingDecompositions}%
  \BibitemOpen
  \bibfield  {author} {\bibinfo {author} {\bibfnamefont {A.}~\bibnamefont
  {Kissinger}}\ and\ \bibinfo {author} {\bibfnamefont {J.}~\bibnamefont {van~de
  Wetering}},\ }\bibfield  {title} {\bibinfo {title} {{Simulating quantum
  circuits with ZX-calculus reduced stabiliser decompositions}},\ }\href
  {https://doi.org/10.1088/2058-9565/ac5d20} {\bibfield  {journal} {\bibinfo
  {journal} {Quantum Science and Technology}\ }\textbf {\bibinfo {volume}
  {7}},\ \bibinfo {pages} {044001} (\bibinfo {year} {2022})}\BibitemShut
  {NoStop}%
\bibitem [{\citenamefont {Nita}\ \emph {et~al.}(2023)\citenamefont {Nita},
  \citenamefont {Mazzoli~Smith}, \citenamefont {Chancellor},\ and\
  \citenamefont {Cramman}}]{Nita2023TheProblem-solving}%
  \BibitemOpen
  \bibfield  {author} {\bibinfo {author} {\bibfnamefont {L.}~\bibnamefont
  {Nita}}, \bibinfo {author} {\bibfnamefont {L.}~\bibnamefont {Mazzoli~Smith}},
  \bibinfo {author} {\bibfnamefont {N.}~\bibnamefont {Chancellor}},\ and\
  \bibinfo {author} {\bibfnamefont {H.}~\bibnamefont {Cramman}},\ }\bibfield
  {title} {\bibinfo {title} {{The challenge and opportunities of quantum
  literacy for future education and transdisciplinary problem-solving}},\
  }\href {https://doi.org/10.1080/02635143.2021.1920905} {\bibfield  {journal}
  {\bibinfo  {journal} {Research in Science {\&} Technological Education}\
  }\textbf {\bibinfo {volume} {41}},\ \bibinfo {pages} {564} (\bibinfo {year}
  {2023})}\BibitemShut {NoStop}%
\bibitem [{\citenamefont {Kurzy{\'{n}}ski}\ \emph {et~al.}(2016)\citenamefont
  {Kurzy{\'{n}}ski}, \citenamefont {Ko{\l}odziejski}, \citenamefont
  {Laskowski},\ and\ \citenamefont
  {Markiewicz}}]{Kurzynski2016Three-dimensionalQutrit}%
  \BibitemOpen
  \bibfield  {author} {\bibinfo {author} {\bibfnamefont {P.}~\bibnamefont
  {Kurzy{\'{n}}ski}}, \bibinfo {author} {\bibfnamefont {A.}~\bibnamefont
  {Ko{\l}odziejski}}, \bibinfo {author} {\bibfnamefont {W.}~\bibnamefont
  {Laskowski}},\ and\ \bibinfo {author} {\bibfnamefont {M.}~\bibnamefont
  {Markiewicz}},\ }\bibfield  {title} {\bibinfo {title} {{Three-dimensional
  visualization of a qutrit}},\ }\href
  {https://doi.org/10.1103/PhysRevA.93.062126} {\bibfield  {journal} {\bibinfo
  {journal} {Physical Review A}\ }\textbf {\bibinfo {volume} {93}},\ \bibinfo
  {pages} {062126} (\bibinfo {year} {2016})}\BibitemShut {NoStop}%
\bibitem [{\citenamefont {Jevtic}\ \emph {et~al.}(2014)\citenamefont {Jevtic},
  \citenamefont {Pusey}, \citenamefont {Jennings},\ and\ \citenamefont
  {Rudolph}}]{Jevtic2014QuantumEllipsoids}%
  \BibitemOpen
  \bibfield  {author} {\bibinfo {author} {\bibfnamefont {S.}~\bibnamefont
  {Jevtic}}, \bibinfo {author} {\bibfnamefont {M.}~\bibnamefont {Pusey}},
  \bibinfo {author} {\bibfnamefont {D.}~\bibnamefont {Jennings}},\ and\
  \bibinfo {author} {\bibfnamefont {T.}~\bibnamefont {Rudolph}},\ }\bibfield
  {title} {\bibinfo {title} {{Quantum Steering Ellipsoids}},\ }\href
  {https://doi.org/10.1103/PhysRevLett.113.020402} {\bibfield  {journal}
  {\bibinfo  {journal} {Physical Review Letters}\ }\textbf {\bibinfo {volume}
  {113}},\ \bibinfo {pages} {020402} (\bibinfo {year} {2014})}\BibitemShut
  {NoStop}%
\bibitem [{\citenamefont {Gokhale}\ \emph {et~al.}(2019)\citenamefont
  {Gokhale}, \citenamefont {Baker}, \citenamefont {Duckering}, \citenamefont
  {Brown}, \citenamefont {Brown},\ and\ \citenamefont {Chong}}]{Gokhale2019}%
  \BibitemOpen
  \bibfield  {author} {\bibinfo {author} {\bibfnamefont {P.}~\bibnamefont
  {Gokhale}}, \bibinfo {author} {\bibfnamefont {J.~M.}\ \bibnamefont {Baker}},
  \bibinfo {author} {\bibfnamefont {C.}~\bibnamefont {Duckering}}, \bibinfo
  {author} {\bibfnamefont {N.~C.}\ \bibnamefont {Brown}}, \bibinfo {author}
  {\bibfnamefont {K.~R.}\ \bibnamefont {Brown}},\ and\ \bibinfo {author}
  {\bibfnamefont {F.~T.}\ \bibnamefont {Chong}},\ }\bibfield  {title} {\bibinfo
  {title} {{Asymptotic improvements to quantum circuits via qutrits}},\ }in\
  \href {https://doi.org/10.1145/3307650.3322253} {\emph {\bibinfo {booktitle}
  {Proceedings of the 46th International Symposium on Computer Architecture}}}\
  (\bibinfo  {publisher} {ACM},\ \bibinfo {address} {New York, NY, USA},\
  \bibinfo {year} {2019})\ pp.\ \bibinfo {pages} {554--566}\BibitemShut
  {NoStop}%
\bibitem [{\citenamefont {Durt}\ \emph {et~al.}(2003)\citenamefont {Durt},
  \citenamefont {Cerf}, \citenamefont {Gisin},\ and\ \citenamefont
  {{\.{Z}}ukowski}}]{Durt2003SecurityQutrits}%
  \BibitemOpen
  \bibfield  {author} {\bibinfo {author} {\bibfnamefont {T.}~\bibnamefont
  {Durt}}, \bibinfo {author} {\bibfnamefont {N.~J.}\ \bibnamefont {Cerf}},
  \bibinfo {author} {\bibfnamefont {N.}~\bibnamefont {Gisin}},\ and\ \bibinfo
  {author} {\bibfnamefont {M.}~\bibnamefont {{\.{Z}}ukowski}},\ }\bibfield
  {title} {\bibinfo {title} {{Security of quantum key distribution with
  entangled qutrits}},\ }\href {https://doi.org/10.1103/PhysRevA.67.012311}
  {\bibfield  {journal} {\bibinfo  {journal} {Physical Review A}\ }\textbf
  {\bibinfo {volume} {67}},\ \bibinfo {pages} {6} (\bibinfo {year}
  {2003})}\BibitemShut {NoStop}%
\bibitem [{\citenamefont {Liu}\ \emph {et~al.}(2021)\citenamefont {Liu},
  \citenamefont {Sun}, \citenamefont {Pachos}, \citenamefont {Yang},
  \citenamefont {Meng}, \citenamefont {Liao}, \citenamefont {Li}, \citenamefont
  {Wang}, \citenamefont {Luo}, \citenamefont {He}, \citenamefont {Huang},
  \citenamefont {Ding}, \citenamefont {Xu}, \citenamefont {Han}, \citenamefont
  {Li},\ and\ \citenamefont {Guo}}]{Liu2021TopologicalParafermions}%
  \BibitemOpen
  \bibfield  {author} {\bibinfo {author} {\bibfnamefont {Z.-H.}\ \bibnamefont
  {Liu}}, \bibinfo {author} {\bibfnamefont {K.}~\bibnamefont {Sun}}, \bibinfo
  {author} {\bibfnamefont {J.~K.}\ \bibnamefont {Pachos}}, \bibinfo {author}
  {\bibfnamefont {M.}~\bibnamefont {Yang}}, \bibinfo {author} {\bibfnamefont
  {Y.}~\bibnamefont {Meng}}, \bibinfo {author} {\bibfnamefont {Y.-W.}\
  \bibnamefont {Liao}}, \bibinfo {author} {\bibfnamefont {Q.}~\bibnamefont
  {Li}}, \bibinfo {author} {\bibfnamefont {J.-F.}\ \bibnamefont {Wang}},
  \bibinfo {author} {\bibfnamefont {Z.-Y.}\ \bibnamefont {Luo}}, \bibinfo
  {author} {\bibfnamefont {Y.-F.}\ \bibnamefont {He}}, \bibinfo {author}
  {\bibfnamefont {D.-Y.}\ \bibnamefont {Huang}}, \bibinfo {author}
  {\bibfnamefont {G.-R.}\ \bibnamefont {Ding}}, \bibinfo {author}
  {\bibfnamefont {J.-S.}\ \bibnamefont {Xu}}, \bibinfo {author} {\bibfnamefont
  {Y.-J.}\ \bibnamefont {Han}}, \bibinfo {author} {\bibfnamefont {C.-F.}\
  \bibnamefont {Li}},\ and\ \bibinfo {author} {\bibfnamefont {G.-C.}\
  \bibnamefont {Guo}},\ }\bibfield  {title} {\bibinfo {title} {{Topological
  Contextuality and Anyonic Statistics of Photonic-Encoded Parafermions}},\
  }\href {https://doi.org/10.1103/PRXQuantum.2.030323} {\bibfield  {journal}
  {\bibinfo  {journal} {PRX Quantum}\ }\textbf {\bibinfo {volume} {2}},\
  \bibinfo {pages} {030323} (\bibinfo {year} {2021})}\BibitemShut {NoStop}%
\bibitem [{\citenamefont {Duan}\ \emph {et~al.}(2001)\citenamefont {Duan},
  \citenamefont {Lukin}, \citenamefont {Cirac},\ and\ \citenamefont
  {Zoller}}]{Duan2001Long-distanceOptics}%
  \BibitemOpen
  \bibfield  {author} {\bibinfo {author} {\bibfnamefont {L.-m.}\ \bibnamefont
  {Duan}}, \bibinfo {author} {\bibfnamefont {M.~D.}\ \bibnamefont {Lukin}},
  \bibinfo {author} {\bibfnamefont {J.~I.}\ \bibnamefont {Cirac}},\ and\
  \bibinfo {author} {\bibfnamefont {P.}~\bibnamefont {Zoller}},\ }\bibfield
  {title} {\bibinfo {title} {{Long-distance quantum communication with atomic
  ensembles and linear optics}},\ }\href {www.nature.com} {\bibfield  {journal}
  {\bibinfo  {journal} {Nature}\ }\textbf {\bibinfo {volume} {414}} (\bibinfo
  {year} {2001})}\BibitemShut {NoStop}%
\bibitem [{\citenamefont {Verresen}(2023)}]{Verresen2023EverythingModel}%
  \BibitemOpen
  \bibfield  {author} {\bibinfo {author} {\bibfnamefont {R.}~\bibnamefont
  {Verresen}},\ }\bibfield  {title} {\bibinfo {title} {{Everything is a quantum
  Ising model}},\ }\href {http://arxiv.org/abs/2301.11917} {\bibfield
  {journal} {\bibinfo  {journal} {ArXiv}\ } (\bibinfo {year}
  {2023})}\BibitemShut {NoStop}%
\bibitem [{\citenamefont {Benhemou}\ \emph {et~al.}(2021)\citenamefont
  {Benhemou}, \citenamefont {Angkhanawin}, \citenamefont {Adams}, \citenamefont
  {Browne},\ and\ \citenamefont {Pachos}}]{Benhemou2021UniversalityAtoms}%
  \BibitemOpen
  \bibfield  {author} {\bibinfo {author} {\bibfnamefont {A.}~\bibnamefont
  {Benhemou}}, \bibinfo {author} {\bibfnamefont {T.}~\bibnamefont
  {Angkhanawin}}, \bibinfo {author} {\bibfnamefont {C.~S.}\ \bibnamefont
  {Adams}}, \bibinfo {author} {\bibfnamefont {D.~E.}\ \bibnamefont {Browne}},\
  and\ \bibinfo {author} {\bibfnamefont {J.~K.}\ \bibnamefont {Pachos}},\
  }\bibfield  {title} {\bibinfo {title} {{Universality of Z3 parafermions via
  edge mode interaction and quantum simulation of topological space evolution
  with Rydberg atoms}},\ }\href {http://arxiv.org/abs/2111.04132} {\bibfield
  {journal} {\bibinfo  {journal} {ArXiv}\ } (\bibinfo {year}
  {2021})}\BibitemShut {NoStop}%
\bibitem [{\citenamefont {Bertlmann}\ and\ \citenamefont
  {Krammer}(2008)}]{Bertlmann2008BlochQudits}%
  \BibitemOpen
  \bibfield  {author} {\bibinfo {author} {\bibfnamefont {R.~A.}\ \bibnamefont
  {Bertlmann}}\ and\ \bibinfo {author} {\bibfnamefont {P.}~\bibnamefont
  {Krammer}},\ }\bibfield  {title} {\bibinfo {title} {{Bloch vectors for
  qudits}},\ }\href {https://doi.org/10.1088/1751-8113/41/23/235303} {\bibfield
   {journal} {\bibinfo  {journal} {Journal of Physics A: Mathematical and
  Theoretical}\ }\textbf {\bibinfo {volume} {41}},\ \bibinfo {pages} {235303}
  (\bibinfo {year} {2008})}\BibitemShut {NoStop}%
\bibitem [{\citenamefont {Gea-Banacloche}\ \emph {et~al.}(1995)\citenamefont
  {Gea-Banacloche}, \citenamefont {Li}, \citenamefont {Jin},\ and\
  \citenamefont {Xiao}}]{Gea-Banacloche1995ElectromagneticallyExperiment}%
  \BibitemOpen
  \bibfield  {author} {\bibinfo {author} {\bibfnamefont {J.}~\bibnamefont
  {Gea-Banacloche}}, \bibinfo {author} {\bibfnamefont {Y.-q.}\ \bibnamefont
  {Li}}, \bibinfo {author} {\bibfnamefont {S.-z.}\ \bibnamefont {Jin}},\ and\
  \bibinfo {author} {\bibfnamefont {M.}~\bibnamefont {Xiao}},\ }\bibfield
  {title} {\bibinfo {title} {{Electromagnetically induced transparency in
  ladder-type inhomogeneously broadened media: Theory and experiment}},\ }\href
  {https://doi.org/10.1103/PhysRevA.51.576} {\bibfield  {journal} {\bibinfo
  {journal} {Physical Review A}\ }\textbf {\bibinfo {volume} {51}},\ \bibinfo
  {pages} {576} (\bibinfo {year} {1995})}\BibitemShut {NoStop}%
\end{thebibliography}%

\end{document}